Calibration of CNEOS Fireball Velocities and the Robustness of Nominally Hyperbolic Events

**Volkan Duran[1] and Abraham Loeb[2]**

[1] Iğdır University, Türkiye
ORCID: 0000-0003-0692-0265 | volkan.duran@igdir.edu.tr

[2] Astronomy Department, Harvard University, Cambridge, Massachusetts, USA
ORCID: 0000-0003-4330-287X | aloeb@cfa.harvard.edu

**Abstract**

The NASA/JPL CNEOS Fireball Data catalogue is one of the few near-global public records of bright atmospheric entries. Its main limitation for orbit work is that it does not publish a velocity covariance for each event. We ask a simple question: how reliable are the CNEOS events that appear hyperbolic? We calibrate the catalogue with 19 events that have independent reference trajectories. In the nine modern high-quality comparisons, the CNEOS-minus-reference speed residual has a mean of -1.00 km $s^{-1}$ and a standard deviation of 0.86 km $s^{-1}$. Variance decomposition assigns 0.77 km $s^{-1}$ to the CNEOS component. The radiant residuals are mostly described by a 0.74° per-axis core, with one 16.3° large discrepancy. The rate of rare, large direction errors is therefore not well constrained. The CNEOS speed scale is robust to reference-error assumptions (0.67–0.88 km $s^{-1}$), but probabilities for the strongest candidates are set by the poorly measured tail. Seven of 354 complete CNEOS states are nominally hyperbolic when the published vectors are used as reported. A +0.9935 km $s^{-1}$ shift, used to test a possible atmospheric-stage or velocity-definition offset, adds five crossings. Polar-IM (2026 April 1) is the strongest post-2018 CNEOS anomaly. Independent GLM-LI stereo solutions give 57.3 ± 2.0 and 56.7 ± 3.0 km $s^{-1}$, close to the parabolic thresholds for their measured directions. CNEOS can identify events worth following up, but a secure interstellar classification requires an independent three-dimensional trajectory with quantified uncertainty for the same event. We recommend public release of per-event CNEOS velocity uncertainties.

## 1. Introduction

The discovery of 1I/‘Oumuamua (Meech et al. 2017), 2I/Borisov (Guzik et al. 2020), and 3I/ATLAS (Seligman et al. 2025) shows that macroscopic interstellar bodies pass through the inner Solar System. Bright fireballs could reveal much smaller interstellar objects because Earth's atmosphere provides a large collecting area. The difficulty is measurement error. An apparently hyperbolic orbit does not automatically imply an interstellar origin (Baggaley 2000; Hajdukova & Kornos 2020; Hajdukova et al. 2024). The CNEOS fireball catalogue, based on U.S. Government sensors, is the only near-global public record of metre-scale atmospheric impactors and includes three-component Earth-fixed velocity vectors for some events (Brown et al. 2002, 2016). It does not publish an uncertainty for each velocity vector. This is the central obstacle to using the catalogue for orbit work: without a per-event uncertainty, every claim that a CNEOS event is hyperbolic rests on an error model imposed from outside the catalogue. Our aim is to measure that error model as directly as the public data allow, and to state clearly which parts of it are well measured and which are not.

A reconstructed heliocentric orbit is hyperbolic when its speed exceeds the local solar escape speed. For many candidates, the excess is only a few kilometres per second. That is comparable to plausible measurement error, so the catalogue value alone cannot establish an interstellar origin. Siraj & Loeb (2022a, 2022b) proposed the 2014 January 8 and 2017 March 9 events, designated IM1 and IM2, as interstellar meteors on the basis of their CNEOS velocity vectors, and Brown & Borovička (2023) disputed the interstellar interpretation of IM1. Peña-Asensio et al. (2025) made the first empirical calibration by matching eighteen CNEOS events to independent ground observations. They found a lower-discrepancy regime after 2018 and reported $\sigma_v$ = 0.55 km $s^{-1}$, $\sigma_{RA}$ = 1.35° and $\sigma_{Dec}$ = 0.84°. Cloete & Loeb (2026a, 2026b) used those values in $10^6$ Monte-Carlo realizations and reported three post-2018 candidates with $p_{bound}$ < $3\times10^{-6}$. Concerns about CNEOS velocity accuracy predate those studies (Devillepoix et al. 2019). The broader meteor literature also shows how difficult it is to separate a truly hyperbolic orbit from one created by measurement error (Hajduková & Kornoš 2020; Hajduková et al. 2024).

Very small formal probabilities depend strongly on the error model. Eighteen calibration events can constrain the core of the error distribution reasonably well, but they say much less about rare, large errors. That matters because the calibration record itself contains a large vector discrepancy that a core-only Gaussian model would regard as extremely unlikely. The key question is therefore not only the size of a typical error, but also the frequency of unusual errors. Hajduková et al. (2024) found no evidence for interstellar fireballs in CNEOS at the population level. Any event-level claim should be read in that wider uncertainty context.

Our main question is whether nominally hyperbolic CNEOS events remain convincing after calibration against independent observations. We first separate the residuals by reference quality and observing epoch. We then test whether an approximately 1 km $s^{-1}$ measurement-stage offset changes the classifications. The rare direction-error tail is estimated from direct velocity and radiant comparisons. GLM geolocation and infrasound back azimuth provide separate external checks. Finally, we distinguish a hyperbolic reconstructed orbit from the stronger claim of an interstellar origin. We also identify the conclusions that depend on assumptions the current data cannot test, in particular the frequency of large errors and the behaviour of the error distribution above the calibrated speed range.

Three features distinguish this calibration from earlier work. First, it assigns each reference to a quality class and an observing epoch before fitting, and estimates the CNEOS and reference variances jointly, so that the CNEOS error scale is not inflated by poor references. Second, it treats the core and the tail of the error distribution separately, measures the tail rate directly on the velocity-direction observable, and checks it against GLM positions and infrasound bearings. Third, it reports a model-free statistic, a geometric distance to the parabolic boundary, alongside model-dependent probabilities, and it tests the published candidates against an independent same-event measurement and against an archived earlier version of the catalogue.

The analysis uses a frozen CNEOS snapshot and an expanded 19-event calibration sample with independent orbit information. The published CNEOS vectors are classified first, without correction. We then test the effect of measurement stage and alternative tail models, and we compare same-event observations where they are available. The larger public-data census has a supporting role. Those datasets are not combined into one exposure; each is used only for the quantity it can check. This keeps the main argument focused on the orbital classification. The contribution is a calibrated assessment of which nominally hyperbolic classifications survive specified measurement-error assumptions and where the available evidence ceases to constrain them. This provides a basis for prioritising independent observations and uncertainty-data release, even when the interstellar origin of an individual candidate remains unresolved.

## 2. Data and Samples

### 2.1 The CNEOS sample

The refreshed public CNEOS API contains 1,069 entries. All event dates and twelve public source fields match the preserved 15 August 2026 snapshot used here. Of the full sample, 360 entries include a three-component Earth-fixed velocity vector and 354 have the position, altitude, and vector information needed for orbit determination. Public-vector coverage rises from 163/753 (21.6%) before 2018 to 197/316 (62.3%) from 2018 onward. A logistic model of vector availability gives an odds ratio of 4.15 (95% CI 2.93–5.89) per decade in reported impact energy and 5.68 (4.52–7.14) per elapsed decade. The vector-bearing events are therefore a selected subset, and population statements below refer to that subset. The Fireball Data API v1.2 calls vx, $v_y$ and $v_z$ entry-velocity components, while also saying that returned fields refer to the peak-brightness event. We treat this as a measurement-stage ambiguity and test its effect later.

For 2008 TC3, we apply the published z-component sign correction only in the analysis field and keep the original catalogue value unchanged. Eight of the 357 entries that report both a scalar speed and a velocity vector differ by more than 1% between those two representations. We keep both values rather than forcing them to agree.

### 2.2 Orbit-bearing calibration cohort

The calibration sample contains nineteen events that clearly match an independent numerical orbit or state in a compatible frame. Eighteen come from the Peña-Asensio et al. (2025) benchmark and are recalculated here with a single Earth ephemeris. The nineteenth is the 2023 May 20 Queensland superbolide. Its public release (Silber et al. 2026; Zenodo 20739713) provides the J2000 geocentric radiant and speed directly, so they do not have to be reconstructed from rounded orbital elements. An event is included only when its identity is secure, its reference kinematics are independent, the geocentric frames are compatible, and a numerical reference orbit is available. The observing pipelines themselves do not have to be identical.

We divide the reference data into three classes based on how the kinematics were measured. Class A is pre-impact astrometry (n = 5: 2008 TC3, 2018 LA, 2019 MO, 2022 EB5, 2024 RW1). These objects had heliocentric orbits measured telescopically before entry (Jenniskens et al. 2009; Borovička & Charvát 2009 for 2008 TC3). Their velocity uncertainty at Earth is of order $10^{-2}$ km $s^{-1}$ and is negligible here. Class N is dedicated fireball-network work (n = 9: Košice, Kalabity, Romania, Baird Bay, Flensburg, Novo Mesto, Ådalen, Iberian, Queensland), using EFN, DFN, GMN, WMPL, or multi-network reductions. Published formal errors are typically below 0.1 km $s^{-1}$. Class C is casual or ad-hoc video (n = 5: Buzzard Coulee, Chelyabinsk (Brown et al. 2013), Sarıçiçek, Ozerki, Viñales), where the reference footage is not calibrated to the same standard. Classes A and N together form the high-quality set (HQ, n = 14).

We assign the reference class from the source of the independent measurement before looking at the residual size. In this way, a large disagreement cannot be used after the fact to label a reference as low quality.

### 2.3 Supporting datasets

The paper has one direct calibration sample and several supporting datasets. Table 1 lists them and states what each one is used for. Only the orbit-bearing sample calibrates CNEOS velocities event by event. The larger unmatched catalogues answer narrower questions about quality cuts, source location, measurement stage, or population plausibility. They are not extra CNEOS calibrators. This distinction is important because a large video-meteor catalogue does not directly calibrate a satellite sensor observing a different size and brightness range.

| Product | Source rows / events | Role in this analysis |
|---|---:|---|
| CNEOS API | 1,069 entries; 360 vectors; 354 analysed | Target catalogue; reporting-selection model |
| Orbit-bearing cohort | 19 events (11 from 2018 onward) | Event-level velocity calibration |
| Global Meteor Network | 3,415,540 unique trajectories | Population tails; shower residual controls |
| SonotaCo annual files | 565,902 records | Video-population control; event search |
| EFN824 / EFN179 (Borovička et al. 2022, 2025) | 824 + 179 (18 shared codes) | Atmospheric-stage measurement; precise-orbit control |
| PANSY radar Level 3 | 2,095,187 event IDs | Radar-population tails; covariance audit |
| IAU MDC (CAMSv3/SAAMER/AMOR) | 12,400,435 records | Independent population controls |
| MAARSY head echoes | 1,599,393 IDs | Orbital profile; time/station screening |
| NASA GLM catalogue | 11,451 records; 130 CNEOS matches | Time/location diagnostics; metadata audit |
| FRIPON public release | 1,340 events (1,334 usable) | Formal eccentricity tails |
| EDMOND-hosted releases | 317,830 + 427 + 4,877 | Separate orbit-population controls |
| CAMO/EMCCD release | 386 rows | Full-state and covariance admissibility audit |
| IAU historical catalogues | 908 DMS; 8,916 Hissar; 6,345 photographic | Error-field and publisher-flag audits |
| BLADE | 124 events | Light-curve morphology and reporting selection |
| SOMN–ELFO | 71 meteors; 90 acoustic arrivals | Optical consistency; acoustic geometry |
| Shober et al. Geminid release | 584 events × 4 pipelines | Reported-uncertainty comparability |
| CEDAR / BRAMS releases | 1,604 raw files; 10 paired cases | Source audits; stage-matching demonstration |
| IAU MDC shower database | 1,736 solutions; 946 codes | Shower-anchored calibration test (Section 5.1) |

| Product | Source rows / events | Role in this analysis |
|---|---|---|
| GMN shower-association table | 34,711 nodes; 386 showers | Chance-association control for the same test |
| Interstellar flux-gap constraints | $n_0 = 10^{-4}$ $AU^{-3}$; three published slopes | Astrophysical prior on provenance (Section 5.2) |
| Archived 2019 CNEOS snapshot | 802 events; 208 vectors | Catalogue version stability (Section 4.11) |
| CNEOS–infrasound event database | 362 detections; 137 events | Direction-channel test; impactor properties (Section 5.5) |

Table 1. Main numerical inputs and the role each dataset plays in this study. Overlapping samples are not added together as a single exposure. The final five rows provide additional checks on association, provenance, catalogue stability and source geometry.

# 3. Analysis

## 3.1 Orbital reduction and independent validation

For each event, we combine the reported geodetic position and Earth-fixed velocity into an ITRS state at the event time. We transform this state to the Geocentric Celestial Reference System, including Earth rotation, precession, nutation, and polar motion. Earth's gravity is then removed with a two-body hyperbolic model. The eccentricity and angular-momentum vectors give the incoming asymptote, and the specific orbital energy gives the geocentric hyperbolic excess speed. We next add Earth's heliocentric velocity and evaluate the resulting heliocentric state relative to the local solar-escape boundary. Equation (1) expresses the classification through the signed reciprocal semimajor axis in $AU^{-1}$. Here $r_\oplus$ is Earth's heliocentric distance, $v_\odot$ is the impactor's heliocentric speed, and $\mu_\odot$ is the solar gravitational parameter.

$$u=\left(2/r_\oplus-v_\odot^2/\mu_\odot\right)\cdot AU\,,u<0 \Leftrightarrow v_\odot>v_{esc}=\left(2\,\mu_\odot/r_\oplus\right)^{1/2} \quad (1)$$

We performed the orbital reduction in two independent computational pipelines and compared all 354 complete states used for orbit determination. The median absolute difference is 0.018 km $s^{-1}$ in geocentric excess speed and 0.0028 $AU^{-1}$ in u. Both pipelines identify the same seven nominally unbound states. As a separate check, the heliocentric Earth velocity at the 2026 April 1 epoch is (+5.203, −26.920, −11.668) km $s^{-1}$. This matches the JPL Horizons value quoted by Cloete & Loeb (2026b) to the stated precision. The differences discussed below are therefore not caused by the orbital reduction itself.

## 3.2 Estimating CNEOS and reference error

For calibration event i, define $\Delta v_i = v_{g,CNEOS} - v_{g,reference}$. We model this residual separately by reference class:

$$\Delta v_i=b+\varepsilon_i^{(C)}+\varepsilon_i^{(R,c(i))} \quad (2)$$

$$\epsilon_i^{(C)}\sim N\left(0,\sigma_C^2\right),\epsilon_i^{(R,c)}\sim N\left(0,\sigma_{R,c}^2\right),\sigma_{R,A}=0.02\,km\,s^{-1} \quad (3)$$

Class A provides the cleanest estimate of CNEOS speed error, because its reference uncertainty is very small. We fix the class-A reference variance $\sigma_{R,c}^2$ at (0.02 km $s^{-1})^2$. For class N, we estimate the reference-side variance from the data because both CNEOS and the fireball-network solution contribute to the residual. Separating the two classes lets us estimate those variance components rather than mixing them. Parameters are fitted by maximum likelihood from multiple starting points. Class C remains in the descriptive tables but is not used in the variance decomposition, because an uncalibrated-video reference has no defensible parametric error scale. In Equation (2), the reference-side residual term is the negative of the reference measurement error. The correlation sensitivity in Appendix A.1 defines ρ between the CNEOS and reference measurement errors, so the residual variance is $\sigma_C^2 + \sigma_R^2 - 2\rho\sigma_C\sigma_R$. Unmodelled differences between event classes may also contribute to the fitted reference-side scale.

We treat the direction residual separately. If the two angular components have independent Gaussian errors with per-axis width $\sigma_\theta$, the separation $\Delta\theta$ follows a Rayleigh distribution, with the maximum-likelihood scale given by Equation (4). Equation (5) adds a second component for rare large direction errors. The core has width $\sigma_{core}$ and weight $1 - \pi$; the gross component has width $\sigma_{gross}$ and weight $\pi$. The modern direct sample contains one gross event in nine. That is too little information to estimate $\pi$ precisely, so we use the mean of its Jeffreys posterior Beta(k + ½, n − k + ½) for marginal mean probabilities. Appendix A.2 separately reports how uncertainty in this shared population weight changes P(bound), conditional on the component shapes.

$$\sigma_\theta = \sqrt{\frac{1}{2n}\sum_{i=1}^{n}\left(\Delta\theta_i\right)^2} \tag{4}$$

$$p\left(\Delta v, \Delta\theta\right) = \left(1-\pi\right) N\left(\Delta v; b, \sigma_C{}^2\right) Ray\left(\Delta\theta; \sigma_{core}\right) + \pi\, N\left(\Delta v; b, \sigma_g{}^2\right) Ray\left(\Delta\theta; \sigma_{gross}\right) \tag{5}$$

### 3.3 Possible measurement-stage offset

CNEOS documentation does not clearly specify the measurement stage. The API says that returned fields refer to the peak-brightness event, while the velocity components are labelled as entry velocities. Fireball networks typically report a pre-atmospheric speed. In the EFN824 catalogue (Borovička et al. 2022), 822 events with a valid $V_{max}$ have a median $V_\infty - V_{max}$ of 0.9935 km s⁻¹. The 95th percentile is 4.103 and the maximum is 10.665 km s⁻¹. The median is close to the modern CNEOS-minus-reference offset, suggesting that measurement stage may contribute to the bias. A difference in how CNEOS defines or retrieves its velocity would produce an offset of the same sign; Section 4.3 and Appendix A.4 test the two interpretations against each other. It does not mean that every event needs the same correction. We therefore use the published vectors as reported and apply +0.9935 km s⁻¹, with nearby alternatives, only as a sensitivity test. The official API changelog corrected the label “pre-entry velocity” in March 2024 and changed the component labels to “entry velocity” in July 2025 (NASA/JPL Solar System Dynamics, Fireball Data API v1.2). These terminology changes do not establish a common velocity epoch for all catalogue entries.

### 3.4 Geometric error scales for changing the classification

We also want a result that is not sensitive to the assumed shape of the error tail. For an event with heliocentric speed $v_\odot$ and local escape speed $v_{esc}$, the parabolic boundary is a sphere in geocentric-asymptote velocity space. Its radius is $v_{esc}$ and its centre is at minus Earth's heliocentric velocity. The shortest move to that boundary is the speed margin $|v_\odot - v_{esc}|$ in this space. For comparison in reported-speed units, Equation (6) rescales that exact asymptote-space distance by $v_{\infty,\oplus}/v_{geo}$. This is a first-order scalar proxy; it is not an exact minimisation over the three-dimensional reported vector, whose direction also changes under gravitational deflection:

$$\|\Delta v\|_{min} = \left(v_\odot - v_{esc}\right)\cdot\frac{v_{\infty,\oplus}}{v_{geo}} \tag{6}$$

Equation (6) asks a simple question: what entry-speed-equivalent error scale corresponds to the geometric distance from the boundary? We can compare that value directly with the errors seen in the calibration sample. Equation (7) gives an approximate calibration-error scale by combining the speed residual with the chord implied by the radiant separation. We also report a pure-speed route, found numerically by changing only the reported speed until the heliocentric margin reaches zero. The exact asymptote-vector difference instead has squared magnitude $(v_C - v_R)^2 + 4v_C v_R \sin^2(\Delta\theta/2)$; using one speed for the angular chord in Equation (7) is only an approximation. Neither approximation is used to compute P(bound).

$$\| \Delta v_i \| = \sqrt{(\Delta v_i)^2 + \left[2 v_{g,i} \sin\left(\frac{\Delta \theta_i}{2}\right)\right]^2} \quad (7)$$

### 3.5 Monte Carlo tests

The modern direct sample contains one large radiant discrepancy in nine events, so the extreme tail is poorly measured. We therefore use the Monte Carlo results as sensitivity tests. The calculation uses the modern orbit-bearing speed and radiant core, plus a gross-direction component whose weight is drawn from the Jeffreys posterior of the direct sample. For comparison, we also show the Peña-Asensio/Cloete-Loeb kernel and a several-percent cross-modal tail scale. A numerical P(bound) does not define candidate membership. The primary event-level information is the nominal orbital state, the model-free error needed to reach the parabolic boundary, and any independent observation of the same event. P(bound) is a perturbation fraction conditional on the nominal state and the chosen kernel, not a posterior probability of interstellar origin.

### 3.6 Catalogue-count test

Catalogue counts provide a broad check, but they do not classify individual events. We run the whole-catalogue null simulation under the same atmospheric-stage scenario used for the twelve-state sensitivity ladder. Because that scenario differs from the seven-event nominal catalogue, the result is not a formal null test of the nominal set. Its purpose is simpler: to show how strongly the apparent number of unbound events can depend on the assumed error model. Event-level margins and independent observations carry more weight in the interpretation.

## 4. Results

### 4.1 Velocity residuals by reference class and epoch

Table 2 groups the nineteen calibration events by reference class and epoch. When all events are combined, the residual scatter is 3.45 km $s^{-1}$ and four residuals exceed 3 km $s^{-1}$. Those four are either casual-video comparisons or pre-2018 network comparisons. The pooled tail therefore does not represent a single stationary modern CNEOS error distribution. We cannot attribute every large residual to the reference measurement, but the pattern is enough to justify separating reference quality and epoch.

| Stratum | n | bias (km $s^{-1}$) | scatter (km $s^{-1}$) | RMS (km $s^{-1}$) | max \|Δv\| | median Δθ (°) | max Δθ (°) |
|---|---|---|---|---|---|---|---|
| all19 | 19 | 0.115 | 3.447 | 3.357 | 8.049 | 2.06 | 71.21 |
| modern11 | 11 | -0.938 | 0.781 | 1.197 | 2.146 | 1.27 | 16.31 |
| pre2018_8 | 8 | 1.563 | 5.055 | 4.980 | 8.049 | 6.24 | 71.21 |
| A_astrom | 5 | -0.891 | 1.209 | 1.401 | 2.146 | 1.27 | 16.31 |
| N_network | 9 | 0.883 | 3.091 | 3.045 | 8.049 | 1.30 | 15.59 |
| C_casual | 5 | -0.259 | 5.490 | 4.917 | 8.000 | 6.82 | 71.21 |
| HQ_all | 14 | 0.249 | 2.666 | 2.581 | 8.049 | 1.28 | 16.31 |
| HQ_modern | 9 | -1.001 | 0.857 | 1.286 | 2.146 | 1.02 | 16.31 |
| HQ_pre2018 | 5 | 2.500 | 3.433 | 3.959 | 8.049 | 4.59 | 15.59 |
| A_modern | 4 | -1.329 | 0.817 | 1.506 | 2.146 | 0.93 | 16.31 |
| N_modern | 5 | -0.739 | 0.879 | 1.079 | 1.579 | 1.02 | 1.63 |

Table 2. The nineteen-event calibration sample separated by epoch and reference class. The largest pooled residuals occur in casual-video comparisons or pre-2018 network events. The modern high-quality subset is therefore used for the primary calibration.

A variance-components fit to the high-quality sample across all epochs (n = 14) gives a CNEOS speed component of 1.13 km $s^{-1}$ and a network-reference component of 3.06 km $s^{-1}$. These values describe this small calibration set; they are not universal instrument constants. For the modern high-quality subset (n = 9), the mean residual is -1.03 km $s^{-1}$, with $\sigma_{CNEOS}$ = 0.77 km $s^{-1}$ and $\sigma_{reference}$ = 0.33 km $s^{-1}$.

In this fit the four class-A events primarily constrain σCNEOS, while the five network events constrain the sum of CNEOS and reference variances. The adopted fit assumes independent errors. Fixing the network-reference scale between 0 and 1.0 km s⁻¹, inflating the astrometric scale, or imposing a correlation of ±0.5 gives σCNEOS = 0.67–0.88 km s⁻¹ (Appendix A.1, Table A1). These are conditional sensitivity scenarios, not a confidence interval. Profiling both the common bias and the network-reference scale gives an approximate 95% likelihood interval of 0.42–1.39 km s⁻¹. Re-estimating both nuisance parameters in every leave-one-out fit gives 0.45–0.85 km s⁻¹, with the minimum when 2018 LA is omitted. The four class-A events alone have a sample standard deviation of 0.82 km s⁻¹ (95% normal-theory interval 0.46–3.05). Thus the central estimate near 0.8 km s⁻¹ is stable across the specified reference-error scenarios, but nine events do not determine it precisely. The Table 4 grid extends to 2.0 km s⁻¹.

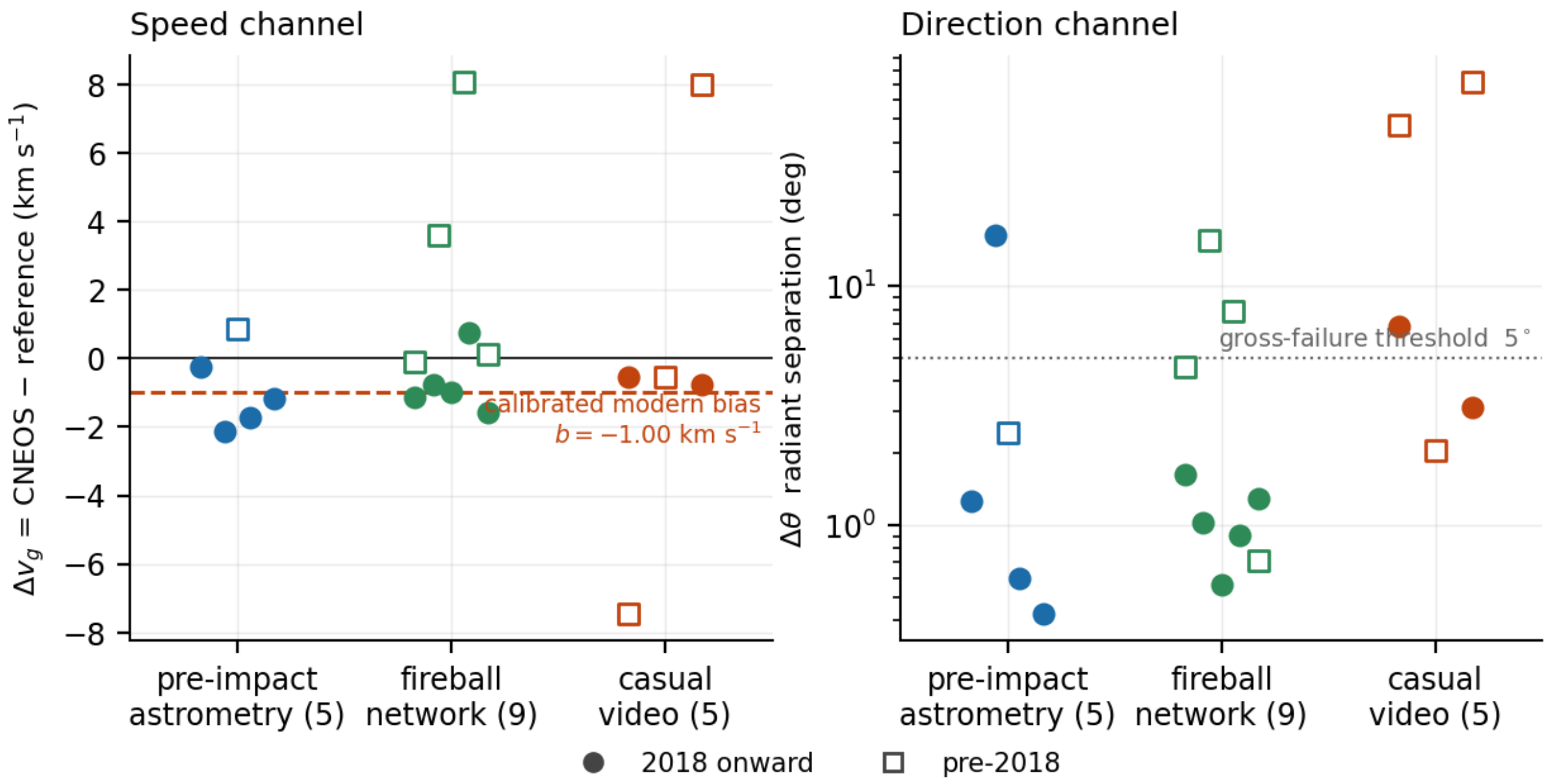

Figure 1. Speed and direction residuals by reference class and epoch. Filled symbols mark events from 2018 onward. The plot motivates using the modern high-quality subset for the primary calibration; it does not assume that every pre-2018 discrepancy comes from the reference measurement.

### 4.2 Modern CNEOS error scale

The primary error model uses the nine modern high-quality events. Their speed residuals have a mean of b = -1.001 km s⁻¹, a standard deviation of 0.857 km s⁻¹, and an RMS about zero of 1.286 km s⁻¹. Eight direction residuals form a core with $\sigma_{\theta,core}$ = 0.736° per axis. The ninth event, 2019 MO, differs by 16.31° from its pre-impact astrometric reference. That discrepancy requires a large-error direction component. With only one such event in nine, however, its frequency remains uncertain. The Jeffreys 95% interval is 1.2-41.4%. We use this direct interval for the rare-event CNEOS-only calculations.

The calibration therefore has two parts that are measured with very different precision. Under the adopted normal and Rayleigh models, the mean speed residual is −1.00 km s⁻¹ (95% interval −1.66 to −0.34 km s⁻¹), the total residual scatter, CNEOS plus reference, is 0.86 km s⁻¹ (0.58–1.64 km s⁻¹), and the angular core is 0.74° per axis (0.55–1.12°). The scatter and the angular core are therefore each known to within a factor of about 1.5–2, and the mean interval excludes zero. These intervals are conditional on the distributional assumptions and, for the angular core, on excluding the large discrepancy; they are not distribution-free guarantees. The tail is not: its frequency is constrained only to 1.2–41%, a range of a factor of 34, and its width rests on a single event. This distinction matters because the leading candidates lie far outside the core. With the core alone, the four strongest nominal events have P(bound) < $10^{-6}$ (Appendix A.2, Table A2). Once the tail is included, more than 95% of their bound probability comes from the gross component, and the same holds for 2009 April 10 (98%) and 2021 May 6 (92%). For the most marginal event, 2015 February 17, the core contributes about 41% and the gross component about 59%. Any probability quoted for the strongest candidates is therefore a statement about the tail, and we treat it as such.

Because a single event, 2019 MO, defines this tail, we repeated the calculation under three interpretations of that event (Table A2). If 2019 MO is treated as an outlier unrelated to CNEOS performance and excluded, the direction model is the 0.74° core alone, and the four strongest candidates have P(bound) < $10^{-6}$. If it is treated as a valid draw from a single error population, one Rayleigh distribution fitted to all nine events has $\sigma_\theta$ = 3.91° per axis; this widens the core but still gives P(bound) ≤ $4\times10^{-5}$ for the four strongest events, because a single Rayleigh law with that width almost never produces errors of 17–42°. If it is treated as a member of a second population, the mixture model gives P(bound) between $2\times10^{-5}$ (2014 January 8) and $1.2\times10^{-2}$ (CNEOS-22), and these values change by factors of about 10 to more than 1000 as the assumed width of the gross component varies from 8° to 24° per axis. The provenance of 2019 MO argues against the first interpretation. Its reference is pre-impact astrometry with negligible error, and the same event also has the largest modern speed residual (−2.15 km $s^{-1}$), which is what a failed CNEOS solution, rather than a mismatched reference, would produce. We therefore keep the second-population model as the primary sensitivity case. The existence of rare large discrepancies does not rest on 2019 MO alone: GLM positions (7 of 125 events) and acoustic bearings (8 of 121 events) show several-percent large discrepancies in other CNEOS-related observables (Sections 4.10 and 5.5), although neither measures the velocity radiant, so they support the presence of a tail but cannot fix its rate or width for the radiant. The spread across the three interpretations is itself the main result: under the absolute speed-error law the four strongest events keep the same order wherever their probabilities can be distinguished, but their numerical P(bound) is not determined by the present data.

The modern speed RMS is larger than the Gaussian width used in recent candidate studies, while the angular core is slightly narrower. The main difference is the 16.3° radiant discrepancy. A core-only Gaussian gives almost no probability to an error of that size. The direct calibration therefore supports a separate large-error direction channel, but the sample is too small to fix its frequency tightly.

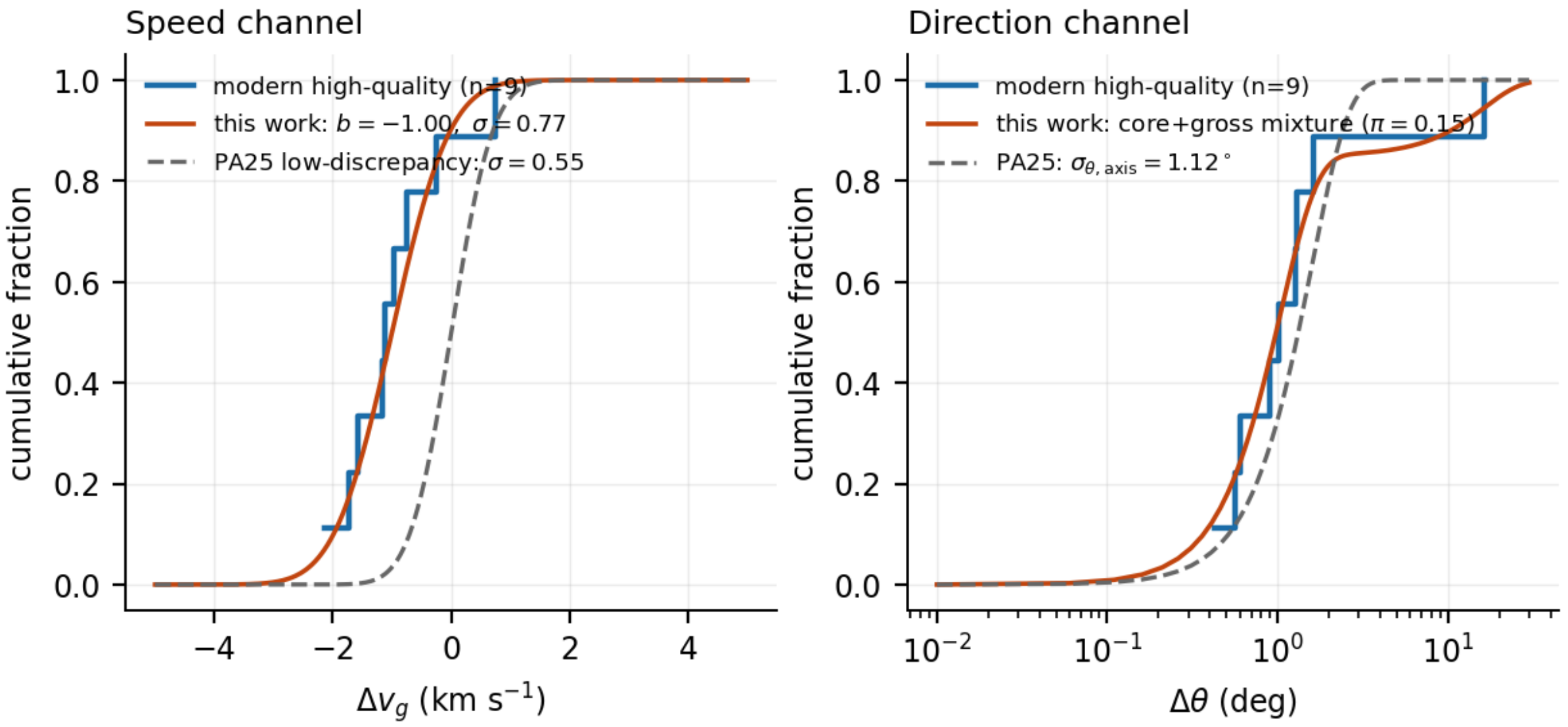

Figure 2. Calibrated error models compared with the nine modern high-quality residuals. Left: speed residuals; a zero-bias σ = 0.55 km $s^{-1}$ Gaussian misses both the mean offset and the observed spread. Right: radiant separations on a logarithmic axis; the calibrated core is narrower than the comparison kernel, while the mixture model can accommodate the observed 16° case.

Within the calibrated speed range, we find no significant trend between residual size and reported speed. For |Δv|, Spearman ρ = +0.29 and p = 0.32 in the high-quality sample; in the modern subset, ρ = −0.23 and p = 0.55. This null result applies only over 6.7–38.6 km $s^{-1}$. As Section 4.8 explains, extrapolating beyond that interval is a more important limitation than the absence of a trend within it.

### 4.3 Measurement-stage sensitivity

The modern mean CNEOS-minus-reference speed residual is -1.001 km s⁻¹. The EFN824 median $V_\infty - V_{max}$ is 0.9935 km s⁻¹ in an independent sample of 822 fireballs. The close agreement in sign and magnitude suggests that measurement stage may explain part of the offset. The EFN distribution is broad, and deceleration depends on altitude, entry geometry, mass, fragmentation, and the definition of the reported velocity. We therefore use the match as evidence that an offset of about this size is present, whatever its cause, not as a universal correction.

The agreement does not by itself distinguish deceleration from a difference in how CNEOS defines or retrieves its velocity, so we examined an illustrative deceleration model (Appendix A.4, Table A4, Figure A1). A single-body drag model calibrated on the EFN824 fireballs has an in-sample association with their measured deceleration (Spearman $\rho$ = 0.68 for 822 events, and 0.84 for the 165 events with maximum brightness below 60 km). Applied to the reported CNEOS heights of the nine modern calibrators, it predicts a mean deceleration of 1.76 km s⁻¹, of the same sign and order of magnitude as the observed −1.00 km s⁻¹ offset. Event by event, however, the predicted deceleration does not track the residual: the regression slope is −0.06 ± 0.22, compared with −1 expected for pure deceleration, and the rank correlation is $\rho = -0.07$ ($p = 0.86$). The only modern calibrator with a positive residual, the 2024 May 18 Iberian event (+0.74 km s⁻¹), is also the one reported highest in the atmosphere (74.3 km), as deceleration would predict, but with nine events the height–residual trend is not significant ($\rho = 0.38$, $p = 0.32$). No additional same-event comparison was found in the two searched EFN releases: none of the 985 distinct fireballs in the two EFN catalogue releases listed in Table 1 (Borovička et al. 2022, 2025) coincides with a CNEOS event, and the closest temporal coincidence is 24.5 s apart but about 7,500 km away. Earlier searches of the larger optical catalogues in Table 1 (GMN, SonotaCo, FRIPON and EDMOND) give the same result: they contain loose coincidences within 300 s and 1,500 km of vector-bearing CNEOS events, but none within 60 s and 300 km, the closest being 41.6 s and 357 km from the 2026 January 30 event. The comparable mean scales make deceleration a plausible contributor, but neither these unmatched samples nor this extrapolated model identifies the cause of the offset. We therefore treat the +0.9935 km s⁻¹ shift as a speed-offset sensitivity test rather than as a deceleration correction. The label "stage sensitivity" is kept in Tables 3 and 6 and Figures 3 and 5 for continuity and should be read in that sense.

The primary classifications use the published CNEOS vectors without a global speed correction. Seven of 354 complete states are then nominally unbound. Applying +0.9935 km s⁻¹ as a sensitivity test raises the count to twelve, with five new crossings. A shift of +1.001 km s⁻¹ also gives twelve, while +1.329 km s⁻¹ gives fourteen. The marginal event count is therefore sensitive to the assumed measurement stage.

The stage issue is especially relevant for high-altitude events. Polar-IM is reported at 90.5 km and the 2026 January 30 event at 89 km. At such altitudes, accumulated deceleration may be smaller than the all-event EFN median. An altitude- and trajectory-resolved model could eventually provide event-specific adjustments. The present data are not sufficient to build one. The two interpretations make different predictions for these events. Under the deceleration interpretation the appropriate shift is much smaller than +0.9935 km s⁻¹: the EFN fireballs with maximum brightness above 90 km have a median deceleration of 0.35 km s⁻¹, and the drag model of Appendix A.4 predicts less than 0.03 km s⁻¹ for metre-scale bodies at 89–90.5 km. Under an illustrative event-independent definition offset, the full shift would apply, and the 2026 January 30 event, whose nominal margin is −0.05 km s⁻¹, becomes nominally unbound. Its classification therefore depends on a question that the public data cannot yet answer.

### 4.4 Nominally hyperbolic events

Table 3 separates the main result from the measurement-stage test. Seven CNEOS states are nominally unbound with the published vectors. Five more cross the parabolic boundary only after the +0.9935 km s⁻¹ sensitivity shift. We keep those five because they are useful follow-up targets, but they are not part of the primary seven-event result. Tail probabilities are shown separately because they depend on the adopted error model; the table instead emphasizes the catalogue state and the model-free error margins.

| Status | Event (UTC) | $E_{imp}$ (kt) | $v_{geo}$ | $v_{\odot}$ nominal | margin nominal | $v_{\odot}$ +0.9935 | margin +0.9935 |
|---|---|---|---|---|---|---|---|
| Nominal | 2014-01-08 17:05:33 | 0.130 | 45.23 | 60.95 | +18.47 | 61.84 | +19.36 |
| Nominal | 2026-04-01 02:13:14 | 0.086 | 69.10 | 51.73 | +9.59 | 52.65 | +10.51 |
| Nominal | 2017-03-09 04:16:37 | 1.100 | 36.70 | 49.81 | +7.53 | 50.64 | +8.36 |
| Nominal | 2022-07-28 01:36:07 | 0.690 | 29.70 | 46.98 | +5.18 | 47.89 | +6.09 |
| Nominal | 2009-04-10 18:42:45 | 0.720 | 19.27 | 44.90 | +2.82 | 46.05 | +3.97 |
| Nominal | 2021-05-06 05:54:26 | 0.180 | 26.81 | 43.51 | +1.58 | 44.36 | +2.43 |
| Nominal | 2015-02-17 13:19:50 | 0.140 | 28.74 | 43.37 | +0.99 | 44.15 | +1.78 |
| Stage sensitivity | 2026-01-30 10:25:37 | 0.120 | 70.75 | 42.39 | -0.05 | 43.37 | +0.93 |
| Stage sensitivity | 2022-07-22 00:16:18 | 0.210 | 17.67 | 41.24 | -0.55 | 42.38 | +0.59 |
| Stage sensitivity | 2015-03-11 06:18:59 | 0.230 | 19.92 | 41.69 | -0.58 | 42.68 | +0.41 |
| Stage sensitivity | 2025-06-26 00:37:20 | 0.390 | 19.58 | 40.96 | -0.82 | 42.03 | +0.25 |
| Stage sensitivity | 2013-07-31 03:50:13 | 0.220 | 17.62 | 40.88 | -0.93 | 41.99 | +0.18 |

Table 3. Nominally unbound states and atmospheric-stage sensitivity crossings. The seven rows marked Nominal are unbound when the published CNEOS vectors are used without correction. The five rows marked Stage sensitivity are bound in the primary reduction and cross the parabolic boundary only after a uniform +0.9935 km $s^{-1}$ shift. Tail probabilities are not used to define membership in either group.

Among the seven nominal states, the 2014 January 8 event (IM1) has the largest nominal hyperbolic margin in the catalogue. Polar-IM is the most informative post-2018 high-speed case because its CNEOS vector gives a large nominal hyperbolic margin and an independent stereo measurement exists for the same event. CNEOS-22 (2022 July 28) is also reproduced from the current catalogue. The 2009 April 10 and 2015 February 17 events need smaller speed changes to become bound, so their classifications are more fragile.

The 2026 January 30 New Zealand event is bound in the primary reduction. Its uncorrected heliocentric margin is approximately -0.05 km $s^{-1}$, and it becomes unbound only after the measurement-stage sensitivity shift. Its high geocentric speed, retrograde geometry, and possible ground observations still make it a valuable calibration target. It is not part of the primary interstellar-candidate set.

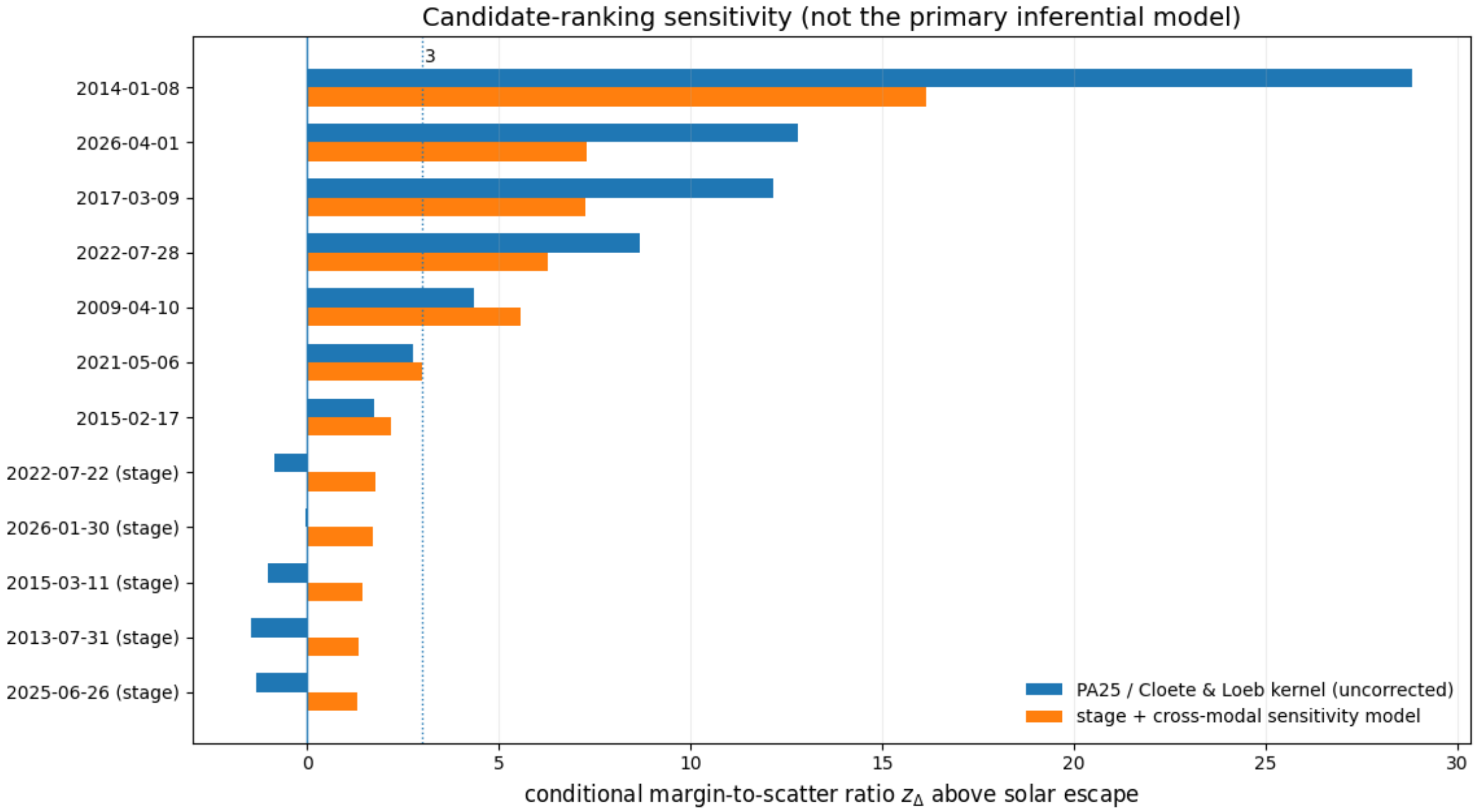

Figure 3. Sensitivity of the candidate ranking. The first series applies the PA25/Cloete & Loeb kernel to the uncorrected vectors; the second uses the atmospheric-stage and cross-modal gross-rate sensitivity construction. The five rows labelled stage are bound in the uncorrected catalogue. The figure is diagnostic and does not define the primary seven-event set.

The 2026 January 30 New Zealand event deserves follow-up for a different reason: it could extend the calibration to higher speed. It is bound in the primary reduction, crosses the boundary only under the stage adjustment, and has a retrograde ecliptic inclination of 168.9°. Public reports suggest that ground observations may exist, but the present dataset contains no independent numerical state with reproducible covariance. A reduced ground trajectory would directly test the stage assumption and extend the modern calibration beyond its current speed range.

Table 4 shows how the ranking of the leading events changes when the speed-error width and gross-direction frequency are varied. The largest-margin events remain near the top, but the numerical P(bound) values change substantially with the assumed gross-direction rate. The ranking is therefore more stable than the probability itself. We use the table to show model sensitivity, not to select a preferred cross-modal failure rate.

| $\sigma_v$ (km s$^{-1}$) | $\pi_{gross}$ | **2014-01-08** | **2026-04-01** | **2017-03-09** | **2022-07-28** | **2009-04-10** |
|---|---|---|---|---|---|---|
| 0.55 | 0.000 | $<10^{-6}$ | $<10^{-6}$ | $<10^{-6}$ | $<10^{-6}$ | $2.5\times10^{-7}$ |
| 0.55 | 0.029 | $5.0\times10^{-6}$ | $4.9\times10^{-4}$ | $1.5\times10^{-3}$ | $2.3\times10^{-3}$ | $4.4\times10^{-4}$ |
| 0.55 | 0.063 | $1.0\times10^{-5}$ | $1.1\times10^{-3}$ | $3.2\times10^{-3}$ | $4.9\times10^{-3}$ | $9.8\times10^{-4}$ |
| 0.55 | 0.110 | $1.6\times10^{-5}$ | $1.8\times10^{-3}$ | $5.6\times10^{-3}$ | $8.6\times10^{-3}$ | $1.7\times10^{-3}$ |
| 0.77 | 0.000 | $<10^{-6}$ | $<10^{-6}$ | $<10^{-6}$ | $<10^{-6}$ | $9.1\times10^{-5}$ |
| 0.77 | 0.029 | $4.8\times10^{-6}$ | $5.1\times10^{-4}$ | $1.5\times10^{-3}$ | $2.3\times10^{-3}$ | $6.9\times10^{-4}$ |
| 0.77 | 0.063 | $1.0\times10^{-5}$ | $1.1\times10^{-3}$ | $3.3\times10^{-3}$ | $5.0\times10^{-3}$ | $1.4\times10^{-3}$ |
| 0.77 | 0.110 | $1.6\times10^{-5}$ | $1.9\times10^{-3}$ | $5.7\times10^{-3}$ | $8.7\times10^{-3}$ | $2.4\times10^{-3}$ |
| 1.29 | 0.000 | $<10^{-6}$ | $<10^{-6}$ | $<10^{-6}$ | $<10^{-6}$ | 0.013 |
| 1.29 | 0.029 | $4.5\times10^{-6}$ | $5.8\times10^{-4}$ | $1.5\times10^{-3}$ | $2.5\times10^{-3}$ | 0.014 |
| 1.29 | 0.063 | $1.0\times10^{-5}$ | $1.3\times10^{-3}$ | $3.4\times10^{-3}$ | $5.3\times10^{-3}$ | 0.015 |
| 1.29 | 0.110 | $1.8\times10^{-5}$ | $2.2\times10^{-3}$ | $5.9\times10^{-3}$ | $9.2\times10^{-3}$ | 0.017 |
| 2.00 | 0.000 | $<10^{-6}$ | $<10^{-6}$ | $1.5\times10^{-6}$ | $5.2\times10^{-4}$ | 0.074 |
| 2.00 | 0.029 | $5.8\times10^{-6}$ | $7.3\times10^{-4}$ | $1.7\times10^{-3}$ | $3.3\times10^{-3}$ | 0.076 |
| 2.00 | 0.063 | $1.2\times10^{-5}$ | $1.6\times10^{-3}$ | $3.7\times10^{-3}$ | $6.4\times10^{-3}$ | 0.077 |
| 2.00 | 0.110 | $2.1\times10^{-5}$ | $2.8\times10^{-3}$ | $6.4\times10^{-3}$ | 0.011 | 0.079 |

Table 4. Conditional P(bound) for the leading events under different speed-width and gross-direction assumptions. These values are CNEOS-only sensitivity checks, not posterior probabilities of interstellar origin. Values are fractions of $4\times10^6$ Monte Carlo draws, computed for the uncorrected vectors with zero-mean speed errors, a 0.74° per-axis direction core and an 11.53° per-axis gross component of weight $\pi_{gross}$; $<10^{-6}$ denotes no bound draw, and a single bound draw corresponds to $2.5\times10^{-7}$.

### 4.5 Catalogue-count test

For the +0.9935 km $s^{-1}$ stage-sensitivity scenario, the conservative bound-only null produces 9.7 apparent unbound states on average, with a 90% range of 5-14. The sensitivity ladder contains twelve. A count of twelve is therefore not unusual under this null. The simulation uses the same global stage shift that created the twelve-state ladder, so it should not be read as a formal null test of the seven-state uncorrected catalogue.

A second construction predicts 3.9 false crossings among events that remain bound after the stage adjustment. The two constructions differ because CNEOS provides no event-level covariance and the true distance of each event from the parabolic boundary is unknown. Catalogue counts alone therefore cannot identify the false-positive rate.

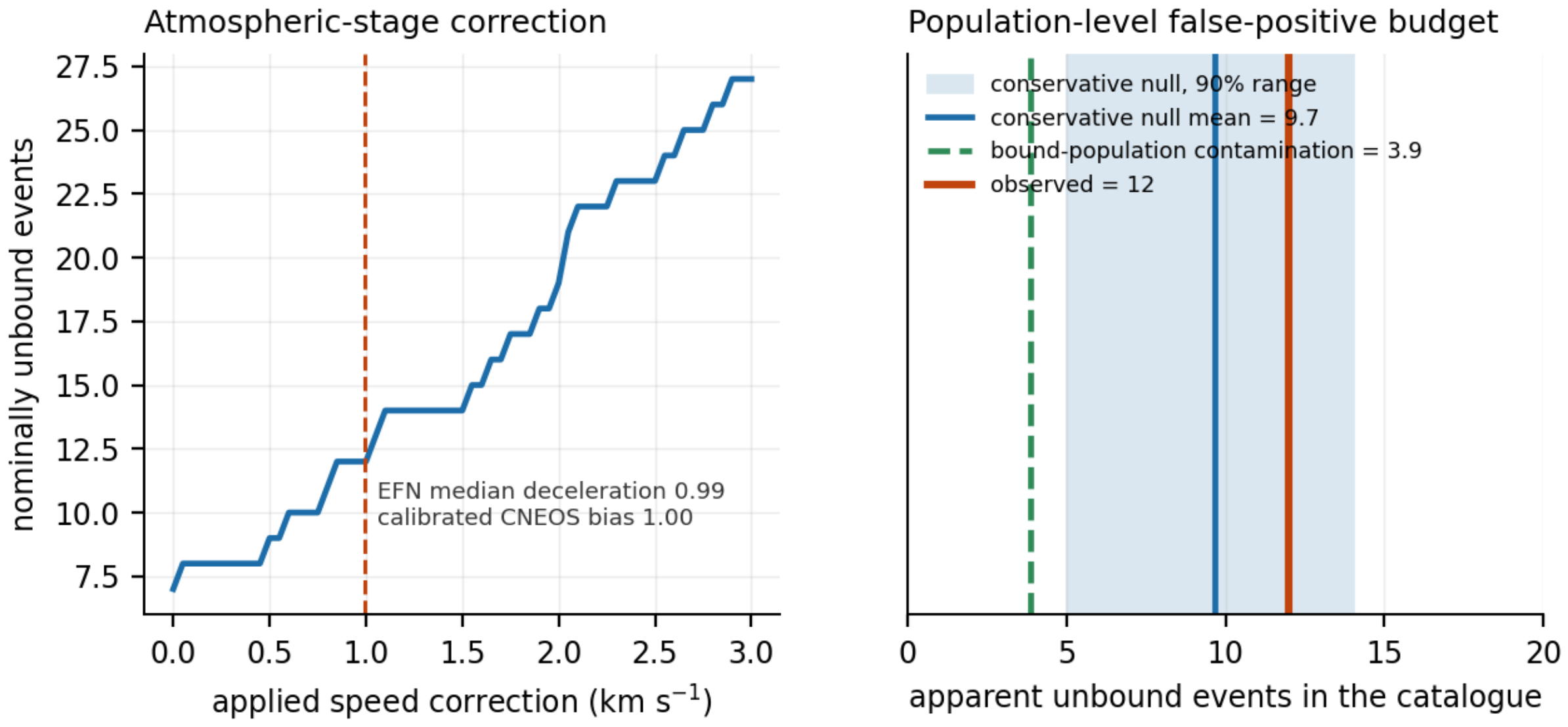

Figure 4. Left: number of nominally unbound states as a function of a global speed shift. Right: expected false-positive count under two null constructions for the same stage-sensitivity scenario. This is a sensitivity analysis, not an estimate of the interstellar population fraction.

Catalogue counts and event-level margins answer different questions. The twelve-state sensitivity count is not unusual under the conservative null. Yet several nominal events require vector changes larger than those seen in the modern direct calibration. That is why they are worth independent measurement. The large required-error margins are a reason to follow up the events, not proof of interstellar origin.

### 4.6 Reproducibility of published candidates

Because the public CNEOS catalogue can change, we checked the published post-2018 candidates against the current API. Polar-IM and CNEOS-22 are reproducible. Our reduction gives heliocentric speeds of 51.73 and 46.98 km $s^{-1}$, matching the published 51.73 and 46.98 km $s^{-1}$.

CNEOS-25 differs under the current public vector. Cloete & Loeb (2026a) report the 2025 February 12 event (73.4° N, 49.3° E, Barents Sea) at $v_\odot$ = 45.63 km $s^{-1}$, exceeding escape by 3.22 ± 0.58 km $s^{-1}$ with pbound < $3\times10^{-6}$. The y-component of the velocity vector for that event was flipped in sign from its original value within a day after the preprint by Cloete & Loeb (2026a) was posted online on the arXiv. The revised vector now reads (+9.5, −19.6, −1.9) km $s^{-1}$, and it yields $v_\odot$ = 37.30 km $s^{-1}$ — bound by 5.10 km $s^{-1}$. Table 5 evaluates all eight sign assignments of the current components: only the two with vy = +19.6 are unbound, and the closest reaches $v_\odot$ = 44.76 km $s^{-1}$, still 0.87 km $s^{-1}$ below the published value.

| $v_x$ | $v_y$ | $v_z$ | $v_\odot$ (km s⁻¹) | $v_{esc}$ (km s⁻¹) | **margin (km s⁻¹)** | **classification** |
|---|---|---|---|---|---|---|
| 9.5 | 19.6 | 1.9 | 43.593 | 42.395 | +1.198 | unbound |
| 9.5 | 19.6 | -1.9 | 44.760 | 42.395 | +2.365 | unbound |
| 9.5 | -19.6 | 1.9 | 36.386 | 42.395 | -6.009 | bound |
| 9.5 | -19.6 | -1.9 | 37.296 | 42.395 | -5.099 | bound |
| -9.5 | 19.6 | 1.9 | 33.379 | 42.395 | -9.016 | bound |
| -9.5 | 19.6 | -1.9 | 34.250 | 42.395 | -8.146 | bound |
| -9.5 | -19.6 | 1.9 | 20.931 | 42.395 | -21.464 | bound |
| -9.5 | -19.6 | -1.9 | 22.132 | 42.395 | -20.263 | bound |

Table 5. All sign assignments of the current public CNEOS vector for 2025 February 12. The first row of each pair uses the component signs in the public CNEOS snapshot analysed here. Cloete & Loeb (2026a) report $v_\odot$ = 45.63 km s⁻¹ for this event, but the posted y-component later changed sign, making the current trajectory bound to the Solar System.

### 4.7 Polar-IM and the independent stereo measurement

Polar-IM lies far from the parabolic boundary in the CNEOS solution. The uncorrected state is 9.59 km s⁻¹ above local solar escape. The first-order entry-speed-equivalent proxy from Equation (6) is 9.47 km s⁻¹, or 13.7% of the reported speed. If only speed is changed, the required shift is 10.56 km s⁻¹. Both values become larger under the +0.9935 km s⁻¹ stage-sensitivity scenario. By the absolute scale of the modern calibration, the CNEOS state is unusual; Section 4.8 discusses how this changes if speed errors scale with speed. The independent stereo measurement is therefore central to the interpretation.

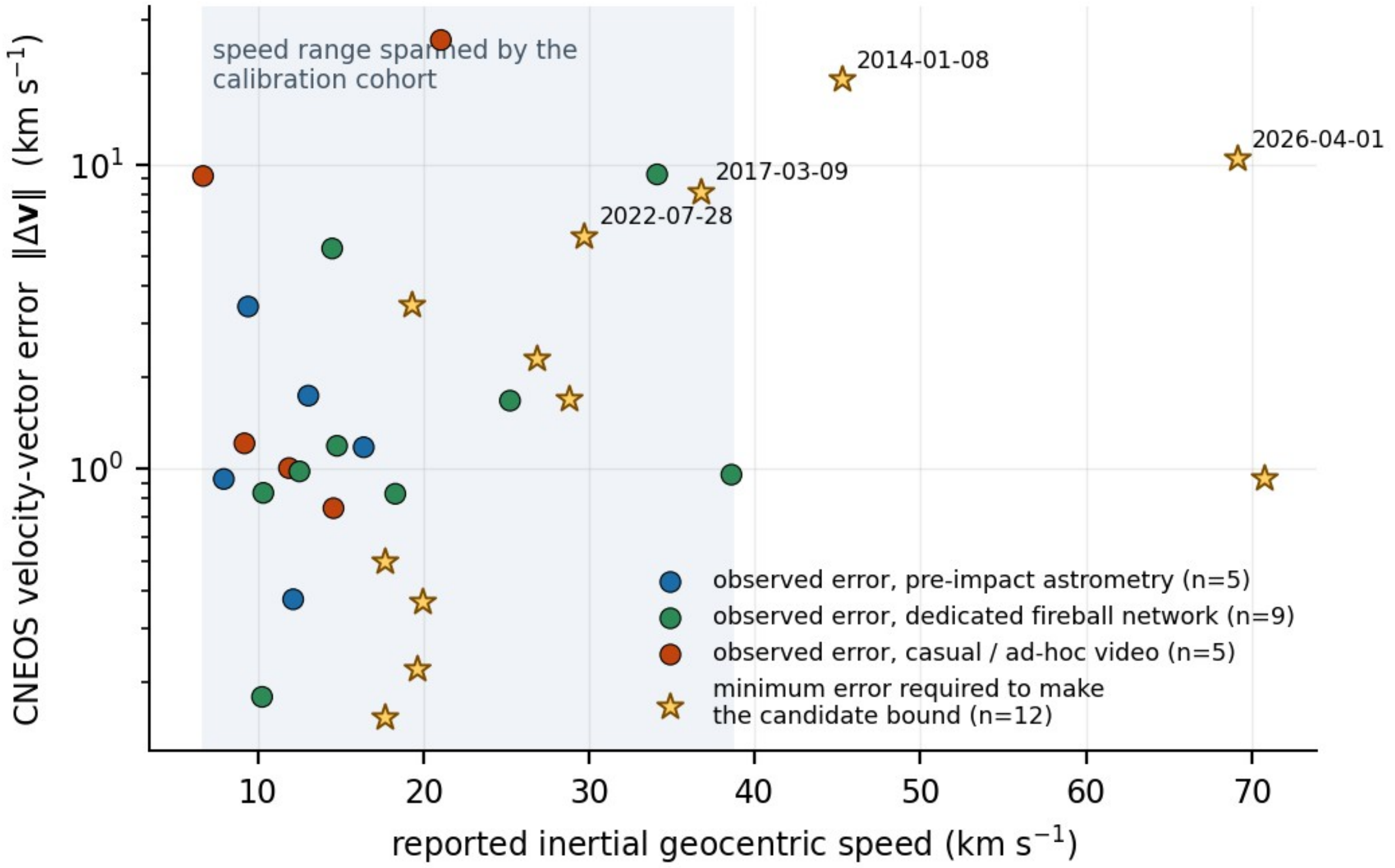

Figure 5. Entry-speed-equivalent geometric error proxies required to move candidate events to the parabolic boundary, compared with the direct calibration sample. Candidate stars use the atmospheric-stage sensitivity ladder from Section 4.3. For the primary uncorrected Polar-IM state, the first-order proxy is 9.47 km s⁻¹. The plot is a scale comparison, not a probability of interstellar origin.

Silber et al. (2026) provide two stereo reductions of the same GOES-GLM and MTG-LI observations. The speeds are 57.3 ± 2.0 and 56.7 ± 3.0 km s⁻¹, with ECEF vectors [3.5, -28.3, 49.7] and [3.5, -27.7, 49.4] km s⁻¹. Their directions are close to the USG/CNEOS direction, but the speeds are about 18% lower. Silber et al. note that this difference can affect the reconstructed orbit. Because both solutions come from the same observations, we treat them as a model average rather than as independent measurements.

A simple speed-only calculation shows the consequence. For the two stereo directions, the parabolic speed thresholds are 58.21 and 58.05 km $s^{-1}$. If the quoted ±2 and ±3 km $s^{-1}$ spreads are treated as Gaussian 1σ scales, each branch places about one-third of its probability above the parabolic boundary. This is only a diagnostic calculation. Silber et al. state that the uncertainties depend on weighting choices and navigation errors, and the full trajectory/state covariance is not public. Without that covariance and a validated model for the large CNEOS speed discrepancy, a combined CNEOS-plus-stereo probability would be model-dependent, so we do not report one.

Speed is only part of the Polar-IM orbit. The stereo direction also enters the reconstruction, and the event lies close enough to the parabolic boundary that covariance among speed, direction, navigation, and deceleration matters. A joint fit to the GLM-LI trajectory with an explicit deceleration model and propagated covariance is therefore the next useful calculation. Until then, Polar-IM remains an unresolved multi-sensor event.

### 4.8 Extrapolation beyond the calibration range

The strongest limitation on the model-based results is the speed range of the direct calibration. The nineteen calibrators span 6.7 to 38.6 km $s^{-1}$, with none above 40 km $s^{-1}$. The two fastest high-priority events are at 45.2 and 69.1 km $s^{-1}$. Their tail probabilities therefore extrapolate the error model beyond the range in which it was directly measured.

We do not know how the error distribution behaves at those higher speeds. At a given mass, faster events may be brighter and easier to detect, which could reduce fractional errors. They are also shorter-lived and may create saturation or geometry problems. No speed dependence is visible within the calibrated range, but the sample has little power to rule out a change above it. The gap can be closed only by common events above 40 km $s^{-1}$ with independent three-dimensional trajectories and quantified uncertainties. Polar-IM is one such comparison. The 2026 January 30 event would be another if a reproducible ground trajectory and covariance were released.

Two illustrative scaling laws test different extrapolations (Appendix A.3, Table A3). If CNEOS speed errors are constant in absolute terms, the modern calibration applies unchanged at all speeds. If they are constant as a fraction of speed, the modern residuals have a fractional standard deviation of 5.8% of the reference vg (a maximum-likelihood fit with a constant bias gives 4.4%), or 3–4 km $s^{-1}$ at vg = 68 km $s^{-1}$. Within the calibrated range the data do not discriminate between the laws. A fractional scatter with a constant bias fits marginally better than the absolute law (ΔAIC = −0.8), a fully fractional law fits slightly worse (ΔAIC = +3.1), and the absolute deviation from the mean bias shows no trend with speed (Spearman $\rho = 0.02$). The nine modern references span only 11.6–37.9 km $s^{-1}$ in vg. These two laws do not bound all possible high-speed error behaviour; speed-dependent bias, correlated speed–direction failures and heavier tails remain uncalibrated.

The choice matters mainly for Polar-IM. Its pure-speed route to the parabolic boundary requires a change of 10.8 km $s^{-1}$ in $v_g$ (10.6 km $s^{-1}$ in reported speed). That is five times the largest modern speed residual in absolute terms (2.15 km $s^{-1}$ for 2019 MO), but in fractional terms it is comparable to that residual: 18.7% of the boundary speed for Polar-IM against 18.6% of the reference speed for 2019 MO, or 15.8% against 22.8% if both are expressed relative to the CNEOS speed. Under the fractional law, the core-only P(bound) of Polar-IM rises from $< 10^{-6}$ to $2\times10^{-4}$–$3\times10^{-3}$ for fractional scales of 4.4–5.8%, and to $6\times10^{-3}$–$1\times10^{-2}$ when the tail is included. By contrast, the 2014 January 8 event requires a 51% speed change or 42° of radiant rotation and remains far outside the calibration under either law, while 2017 March 9 and CNEOS-22, which require 28% and 24%, lie inside the calibrated speed range.

The only same-event comparison above 40 km $s^{-1}$ points in the same direction. The two GLM-LI stereo reductions of Polar-IM are 17–18% slower than the CNEOS speed (21–22% relative to the stereo speeds), a discrepancy comparable to the fractional rather than the absolute scale of the modern residuals. Because the stereo solutions have their own uncertainties, this does not show which measurement is in error. It does mean that the assumption that the absolute calibration measured below 40 km $s^{-1}$ holds at 60–70 km $s^{-1}$ is not supported by the one available test. We therefore regard the CNEOS-only probabilities for Polar-IM as conditional on that assumption.

The direction channel has no high-speed calibration at all. The acoustic-bearing residuals show no trend with reported speed up to 44.8 km $s^{-1}$ (Section 5.6), but they test source location rather than the radiant. Doubling the angular core to 1.47° per axis leaves the four strongest candidates unchanged (core-only P(bound) < $10^{-6}$) because they require radiant changes of 17–42°, but it raises the core-only P(bound) of the marginal events 2021 May 6 and 2015 February 17 from $5\times10^{-3}$ to $1.3\times10^{-2}$ and from 0.05 to 0.09. Within this limited doubling test, the strongest-candidate changes would arise mainly through the tail, which is already the dominant and least constrained term.

### 4.9 Why velocity direction matters

Velocity direction is essential to the classification. At Earth's distance, an object bound to the Sun has $v_\odot \le v_{esc}$. The largest possible geocentric asymptotic speed occurs when the object's heliocentric velocity is opposite to Earth's, giving $v_{\infty,\oplus} \le v_{esc} + v_{Earth}$, approximately 71.95 km $s^{-1}$ near 1 au. None of the reconstructed CNEOS states exceeds this ceiling. A large speed by itself therefore does not prove that an orbit is unbound.

The same geometry gives a direction-based margin. At fixed geocentric asymptotic speed, we solve for the angle to Earth's heliocentric velocity that would place the event on the parabolic boundary. Table 6 gives this calculation for the +0.9935 km $s^{-1}$ stage-sensitivity ladder. For the uncorrected Polar-IM state, the required angular change is about 23.0°. The pure-speed route requires 10.56 km $s^{-1}$. These are two simple ways to see how far the reported state lies from the boundary.

| Event (UTC) | $v_{\infty,\oplus}$ | fraction of ceiling | θ (°) | $\theta_{crit}$ (°) | required radiant error (°) | required speed error (km $s^{-1}$) |
|---|---|---|---|---|---|---|
| 2014-01-08 17:05:33 | 44.86 | 0.617 | 70.8 | 114.5 | 43.69 | 23.07 |
| 2009-04-10 18:42:45 | 16.91 | 0.236 | 18.6 | 53.2 | 34.57 | 3.33 |
| 2026-04-01 02:13:14 | 69.20 | 0.962 | 134.8 | 161.1 | 26.30 | 11.55 |
| 2017-03-09 04:16:37 | 36.00 | 0.498 | 80.2 | 100.9 | 20.74 | 10.43 |
| 2022-07-28 01:36:07 | 28.60 | 0.402 | 68.5 | 87.7 | 19.14 | 6.72 |
| 2021-05-06 05:54:26 | 25.47 | 0.356 | 72.7 | 80.9 | 8.21 | 2.84 |
| 2022-07-22 00:16:18 | 14.96 | 0.210 | 35.6 | 40.9 | 5.23 | 0.52 |
| 2015-02-17 13:19:50 | 27.57 | 0.380 | 80.3 | 85.6 | 5.35 | 2.26 |
| 2026-01-30 10:25:37 | 70.88 | 0.975 | 161.2 | 164.9 | 3.68 | 0.94 |
| 2013-07-31 03:50:13 | 14.90 | 0.209 | 38.9 | 40.5 | 1.55 | 0.16 |
| 2015-03-11 06:18:59 | 17.70 | 0.245 | 55.0 | 57.2 | 2.27 | 0.41 |
| 2025-06-26 00:37:20 | 17.29 | 0.243 | 53.1 | 54.6 | 1.47 | 0.23 |

Table 6. Direction and speed sensitivity for the +0.9935 km $s^{-1}$ stage-adjusted ladder. "Fraction of ceiling" is the reconstructed geocentric asymptotic speed divided by $v_{esc} + v_{Earth}$. The radiant change and the pure-speed change are separate one-parameter routes to the parabolic boundary. For the primary uncorrected Polar-IM state, the corresponding values are approximately 23.0° and 10.56 km $s^{-1}$.

Table 6 shows that, for Polar-IM, both simple routes lie beyond the modern direct calibration in absolute terms; Section 4.8 shows that the speed route is comparable to the largest modern residual if errors scale with speed. In the primary uncorrected state, a radiant rotation of about 23.0° would reach the parabolic boundary, compared with a largest modern radiant discrepancy of 16.31°. Changing only the speed would require 10.56 km $s^{-1}$, compared with a largest modern speed residual of 2.15 km $s^{-1}$. This is why a full three-dimensional uncertainty treatment is more informative than a speed-only model for this event.

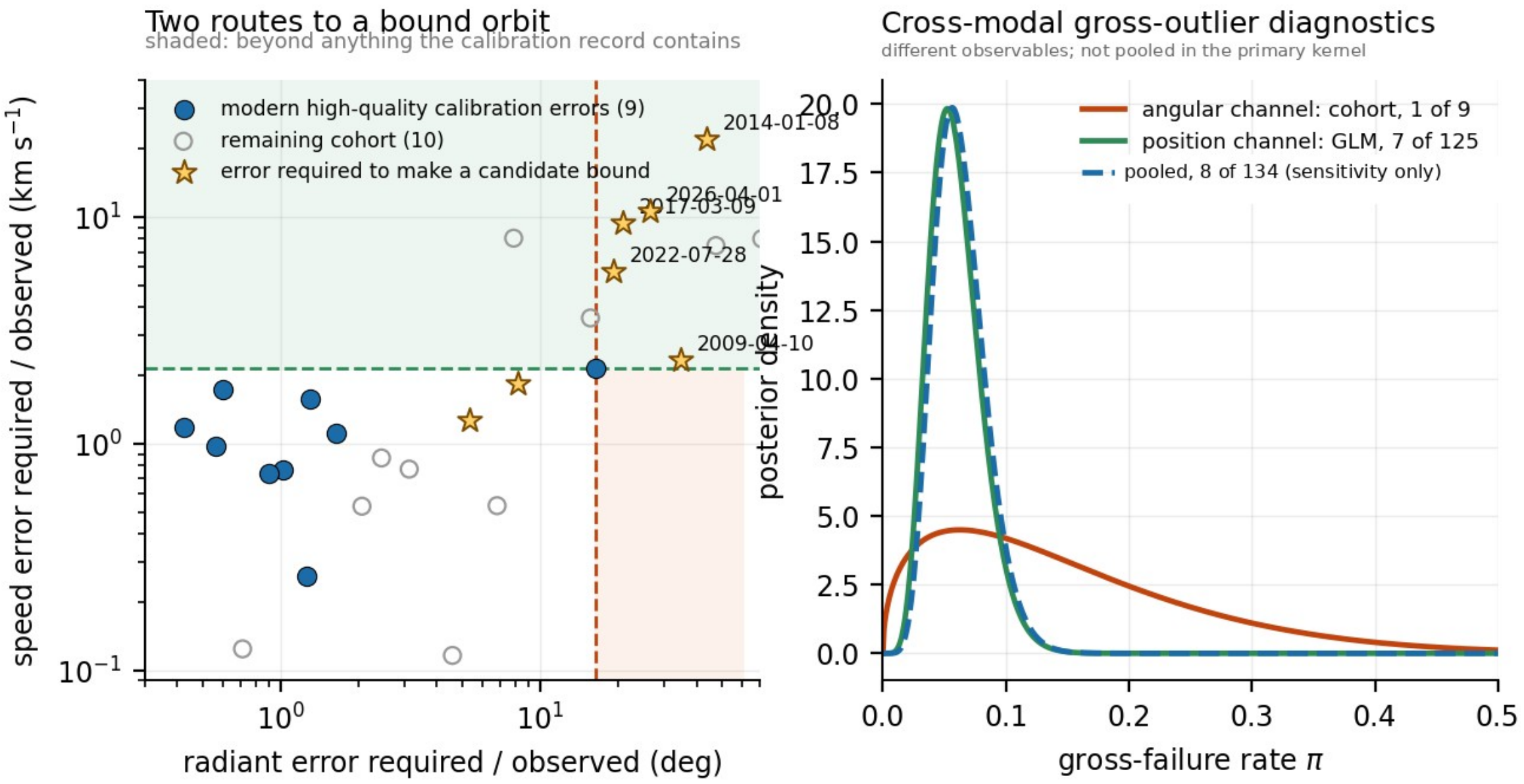

Figure 6. Left: radiant-only and speed-only changes required to move each event to the parabolic boundary, shown against the direct calibration record. Right: Jeffreys posteriors for large discrepancies in the orbit-bearing radiant and GLM position channels. The dashed combined curve is included only as a cross-modal sensitivity illustration and is not used as the primary radiant-error model.

## 4.10 Independent checks on large discrepancies

The direct orbit-bearing sample is too small to measure the frequency of large direction errors precisely. We therefore look at larger datasets for context. Among 125 modern CNEOS-GLM common events with independently measured positions, seven exceed a 100 km position threshold. The corresponding Jeffreys mean is 5.95%, with a 95% interval of 2.54-10.68%. This is a check on large discrepancies in another observable, not a replacement for the direct radiant calibration.

GLM position error and CNEOS radiant error are different quantities. The same is true of the infrasound result in Section 5.5. Its event-level bearing-outlier rate is near 7%, but a back azimuth also contains atmospheric propagation effects and tests source location rather than velocity direction. The similar percentages show that several-percent large discrepancies occur in more than one channel. We do not treat those channels as the same error process.

For comparison, we also evaluate an alternative pooled 6.45% cross-modal calculation and a GLM threshold grid. These show how the numerical tail probabilities move if a several-percent cross-modal rate is used instead of the direct orbit-bearing posterior. The qualitative conclusion for the leading CNEOS-only events is unchanged. The direct model remains the main one because it is calibrated on the velocity-direction observable used in the orbital classification.

## 4.11 Stability of the public CNEOS catalogue

The public CNEOS catalogue changes over time and does not provide a changelog. To measure that instability, we compare the raw 2026 API response with an independently archived scrape covering 1988 to 2019 October. The archive contains 802 events with the same twelve fields, including 208 with velocity components. The comparison also helps interpret the 2025 February 12 discrepancy discussed in Section 4.6.

This comparison uses the untouched API values rather than the analysis fields. The 2008 TC3 sign correction from Section 3.1 therefore does not affect it. The 2026 API still lists $v_z = +3.8$ km s$^{-1}$ for that event, confirming that the correction of Peña-Asensio et al. (2022) has not been incorporated into the public catalogue and must still be applied separately in the analysis.

| Field | Compared | Revised | Rate | Largest change |
|---|---|---|---|---|
| velocity components $v_x$, $v_y$, $v_z$ | 154 | 0 | 0.00% | — |

| Field | Compared | Revised | Rate | Largest change |
|---|---|---|---|---|
| brightness altitude | 294 | 0 | 0.00% | — |
| longitude | 516 | 2 | 0.39% | 345.4° (hemisphere flip) |
| latitude | 516 | 4 | 0.78% | 5.1° |
| calculated total impact energy | 691 | 175 | 25.33% | ×4.53 / ÷3.42 |

Table 7. Field-by-field comparison of the archived 2019 CNEOS snapshot and the 2026 API for 691 events matched by exact timestamp. Velocity vectors and altitudes are unchanged in this historical subset, while calculated impact energies are revised for about a quarter of events.

For the archived pre-2019 subset, none of the 154 comparable velocity vectors and none of the 294 comparable altitudes changed between the 2019 scrape and the 2026 API. That supports reproducibility for this historical subset. It does not establish a zero revision rate for later events. The API documentation allows data-format changes, and CNEOS operational notes describe updates to existing events. Historical stability therefore cannot rule out a later revision to a 2025 event.

Impact energy is much less stable. Of 691 matched events, 175 were revised: 128 upward and 47 downward. The median ratio is 1.073, with extremes of ×4.53 and ÷3.43. Energy enters the reporting-selection model in Section 4.1 and the 0.45 kt low-discrepancy threshold used by Peña-Asensio et al. Between the two catalogue versions, 11 events (1.6%) cross that threshold, so the regime assignment changes for about one event in sixty. Energy also sets the meteoroid radius in the Section 5.2 number-density calculation. Because radius scales with the cube root of energy, the revisions imply a median radius factor of 1.02 and extremes of 0.66 to 1.66. For the most strongly revised events, the required interstellar number density in Table 9 can change by a factor of up to 4.5. We treat this as a systematic uncertainty in the provenance calculation because the current catalogue values are also those used in the recent candidate literature.

Timestamps are also revised. Among 111 events in the 2019 archive with no exact 2026 timestamp match, 107 have a 2026 event within two minutes and are best interpreted as re-timings rather than deletions. Four appear genuinely absent, and 134 events were added within the archived time span. The shifts are small: median 1.0 s, 90th percentile 7.8 s, and maximum 40 s. None reaches the 60-second matching window used here. The common-event searches in Sections 4.5, 6.2 and 7 are therefore unaffected by the level of re-timing seen in this comparison.

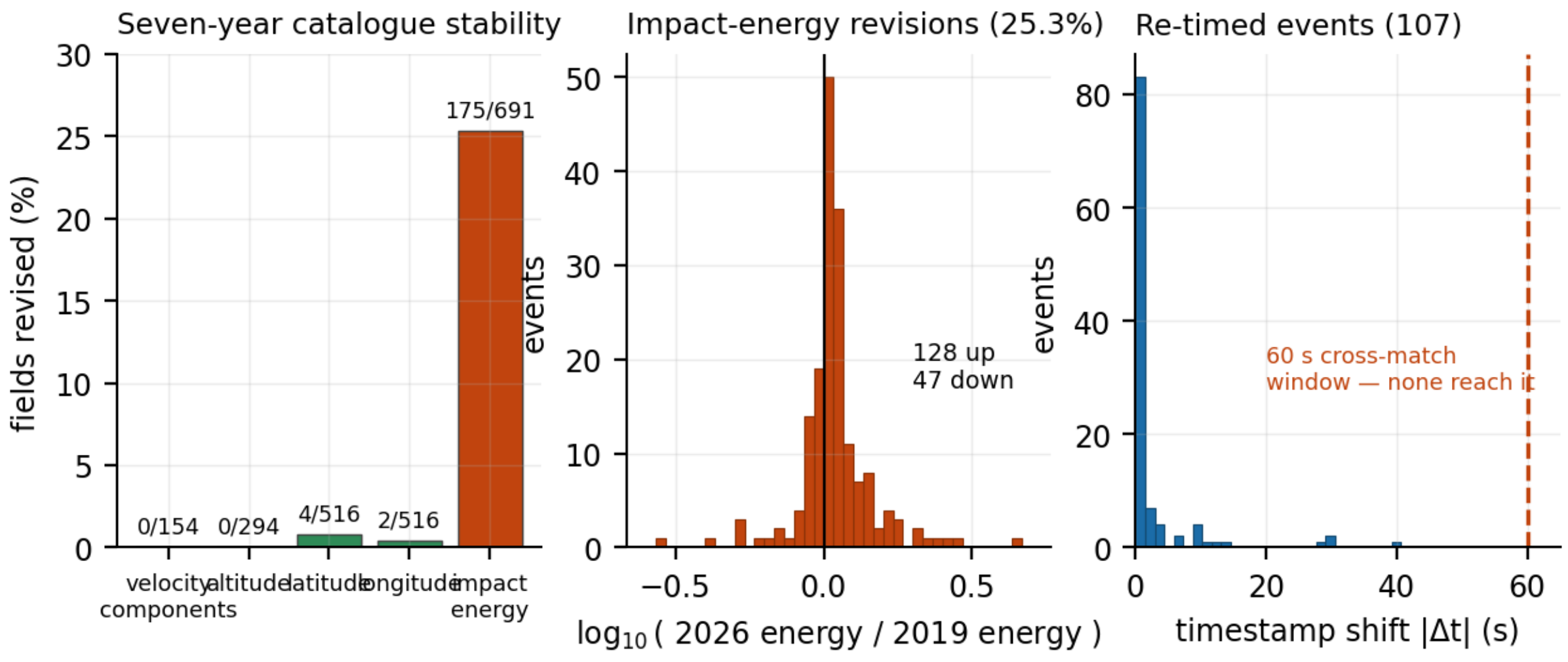

Figure 7. Changes between the archived 2019 CNEOS snapshot and the 2026 API. Left: revision rate by field. Centre: distribution of impact-energy changes. Right: timestamp shifts for re-timed events, compared with the 60-second cross-match window used in this study.

## 5. Independent Tests of the Interstellar Interpretation

The orbit-bearing sample is the direct calibration, but it contains only nineteen events and none is faster than 38.6 km s$^{-1}$. Two follow-up questions can be tested with independent public data. Can another observable extend the kinematic check to higher speed? And is the implied number of interstellar events astrophysically plausible? These tests support the main calibration rather than replace it.

### 5.1 Can meteor showers extend the calibration?

A meteor shower provides an expected pre-atmospheric velocity vector, with some natural stream dispersion. If a CNEOS event can be securely associated with a shower from its time and radiant, the shower speed becomes an external comparison even without an exact common-event trajectory. Major showers also reach higher speeds than most orbit-bearing calibrators. We therefore tested whether secure associations occur often enough to be useful.

We use two public shower products. The IAU Meteor Data Center contains 1,736 published solutions for 946 shower codes, including solar longitude, radiant, radiant drift, and geocentric speed. Requiring at least three independent solutions leaves 176 well-established streams. Their scatter is 0.82 km $s^{-1}$ in speed and 2.76° in right ascension. The Global Meteor Network association table contains 34,711 solar-longitude-resolved nodes for 386 streams, with a median association radius of 3.06°. These products provide an independent test of whether shower anchoring can help the CNEOS calibration.

We select shower membership without using speed. Only solar longitude and radiant direction enter the match, so the speed residual remains an independent outcome. Chance associations are estimated by shifting each event's solar longitude by a random offset while keeping its radiant fixed. This breaks real stream membership while preserving the overall matching geometry.

The shower test finds no excess above chance. A requirement of 6° in solar longitude and 5° from the drift-corrected radiant associates 3 of the 354 vector-bearing CNEOS events. Random solar-longitude shifts give 4.82 ± 2.16 expected associations (P = 0.83). With the full 386-stream GMN table, we find 8 associations, compared with 7.6 ± 2.5 expected. The current sample therefore does not provide enough secure shower matches to extend the high-speed calibration.

This result is consistent with metre-scale CNEOS impactors being mostly sporadic and with fast shower members being rare at these sizes. The single Southern Taurid association has a small speed residual, but one chance-level association cannot calibrate a population. For the public data examined here, independent trajectories of individual fast events remain the more useful route.

### 5.2 Population constraints from interstellar-object surveys

Published interstellar-object constraints provide an independent population check. We use the size-distribution families of Peña-Asensio & Seligman (2025) to ask whether the number of candidate events is compatible with current estimates of the interstellar population. The extrapolation spans many orders of magnitude, from kilometre-scale objects to metre-scale meteoroids, so it is necessarily uncertain. The calculation is therefore a consistency test rather than an event-by-event classification.

Peña-Asensio & Seligman (2025) anchor the kilometre-scale population with 1I/‘Oumuamua and 3I/ATLAS at a number density of about $10^{-4}$ $AU^{-3}$. Reaching meteoroid sizes requires a size-distribution slope. They consider three. The $r^{-3.0}$ slope connects spacecraft dust detections to kilometre-scale objects. The $r^{-2.7}$ slope is the steepest one consistent with CMOR radar limits, where five 3σ hyperbolic candidates appear among eleven million orbits. The $r^{-2.3}$ slope is compatible with optical limits, including the Global Meteor Network result of no secure interstellar meteoroid and an upper limit of about $10^{-6}$ relative to the interplanetary population.

For each candidate, Equation (8) converts reported impact energy and speed into mass and radius at a bulk density ρ = 1500 kg $m^{-3}$. We evaluate the interstellar number density at that radius for each published slope q, then use Equation (9) for the gravitationally focused Earth encounter rate. The kilometre-scale anchor is $n_0 = 10^{-4}$ $AU^{-3}$ at $R_0 = 220$ m. Here $v_\infty$ is the encounter speed and $v_{esc,\oplus}$ is Earth's escape speed. Table 8 gives the sample-level result.

$$m=\frac{2E}{v^2},R=\left(\frac{3m}{4\pi\rho}\right)^{\frac{1}{3}} \quad \textbf{(8)}$$

$$n(>R)=n_0\left(R_0/R\right)^q,\dot{N}=n(>R)\,v_\infty\pi R_\oplus^2\left(1+\frac{v_{esc,\oplus}^2}{v_\infty^2}\right) \quad \textbf{(9)}$$

| Slope | n(>R) at R = 0.3 m ($AU^{-3}$) | Expected interstellar events, 2018-2026 CNEOS vector sample |
|---|---|---|
| $r^{-2.3}$ | 389 | 6.0e-06 |
| $r^{-2.7}$ | 5.45e+03 | 5.2e-05 |
| $r^{-3.0}$ | 3.94e+04 | 2.6e-04 |

Table 8. Interstellar meteoroid abundance at the CNEOS size scale for the three published size-distribution slopes, together with the expected number of genuine interstellar events in the modern vector-bearing sample. Even the steepest slope predicts fewer than $10^{-3}$ events.

All three size-distribution extrapolations predict far fewer than one event. That is a strong population-level tension for an interstellar interpretation of the CNEOS candidates. It does not decide the origin of any single event, because an individual state can still be either genuinely unbound or mismeasured.

### 5.3 Sensitivity to origin assumptions

For completeness, we also use a speed-only Bayes-factor calculation with a calibrated CNEOS speed likelihood and alternative bound and interstellar speed priors. We use it only to test sensitivity to those assumptions. A CNEOS hyperbolic classification depends on both speed and radiant direction, as Section 4.9 shows. A one-dimensional speed Bayes factor is therefore incomplete when the decisive information is in the full three-dimensional state.

$$BF=\frac{\int_U L\left(v_{obs}\mid v\right)\pi_U(v)\,dv}{\int_B L\left(v_{obs}\mid v\right)\pi_B(v)\,dv} \quad \textbf{(10)}$$

In this exploratory calculation, the large-error component prevents an extreme speed residual from producing unlimited evidence for an interstellar origin. The numerical results are useful because they show how strongly the answer depends on the assumed tail. They are not used in the abstract, the candidate ranking, or the conclusions as event-level probabilities of interstellar origin.

The main lesson is simple. Once a non-zero large-error channel is allowed, one catalogue vector cannot generate arbitrarily large evidence ratios. A full event-level Bayes factor would need a joint speed-and-direction likelihood and, for Polar-IM, the independent stereo geometry.

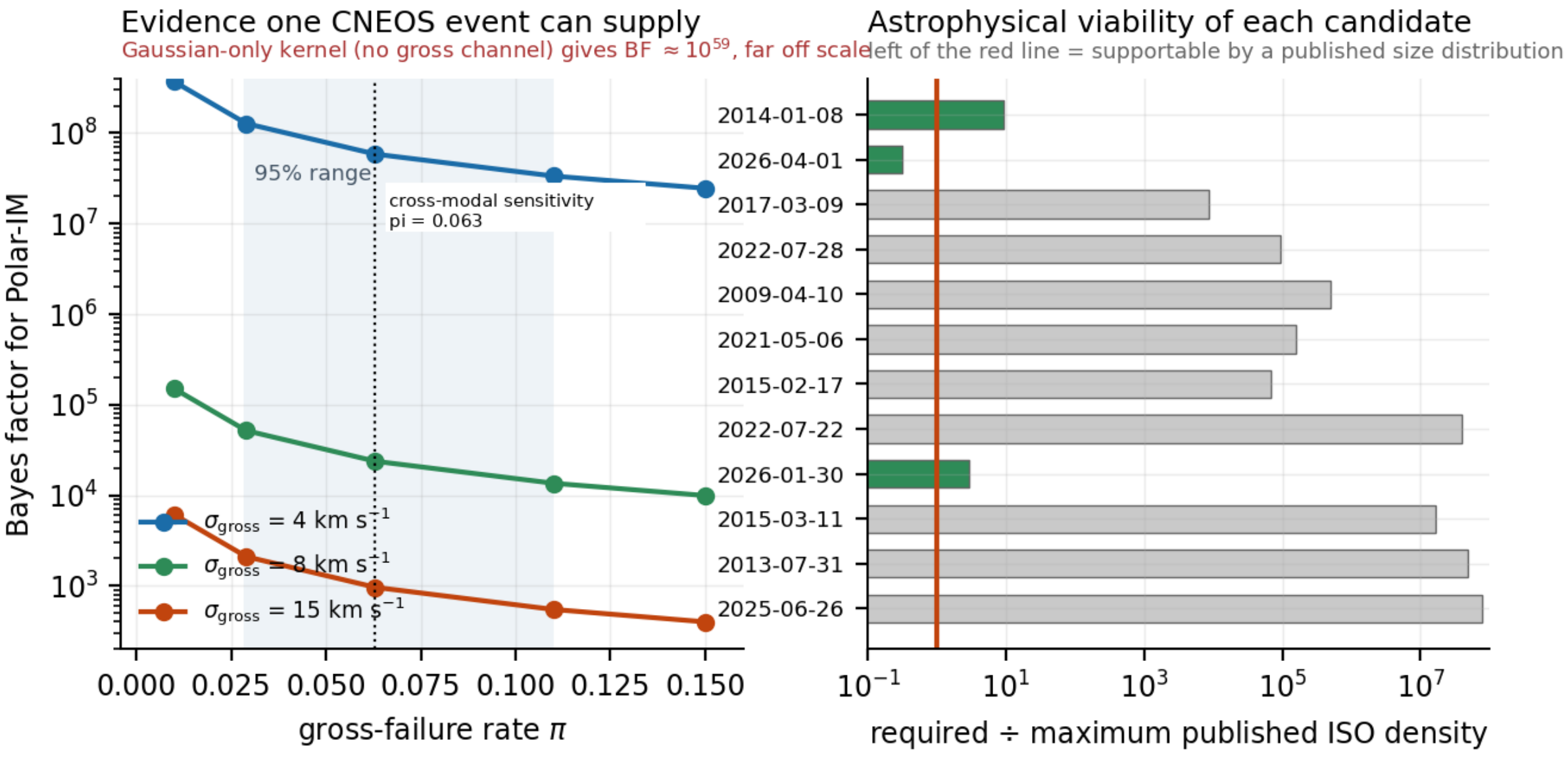

Figure 8. Exploratory sensitivity of the speed-only evidence and the size-distribution consistency calculation. The vertical marker uses the cross-modal gross-outlier value only to show model dependence. It is not the primary radiant-error rate, not a full-state Bayes factor, and not a determination of origin.

### 5.4 Required interstellar density

Table 9 provides an order-of-magnitude check on the origin assumptions. Its most direct quantity is the interstellar number density required to bring the speed-only posterior odds to roughly unity, compared with the density supplied by each published size-distribution extrapolation. The P(IS) columns are model-dependent sensitivity outputs. They are not used as calibrated event-level posterior probabilities or as detection criteria.

$$O_{post} = \frac{N_{IS}}{N_{ev}} \cdot BF \, , n_{req} = \frac{N_{ev}}{\dot{N}_1 T \cdot BF} \quad (11)$$

| Event (UTC) | R (m) | Bayes factor | P(IS) $r^{-2.3}$ | P(IS) $r^{-2.7}$ | P(IS) $r^{-3.0}$ | n required (AU$^{-3}$) | n max published | shortfall |
|---|---|---|---|---|---|---|---|---|
| 2014-01-08 17:05:33 | 0.44 | 2369 | 0.0014 | 0.016 | 0.097 | 1.18e+05 | 1.26e+04 | 9.36 |
| 2026-04-01 02:13:14 | 0.29 | 2.374e+04 | 0.029 | 0.3 | 0.76 | 1.42e+04 | 4.44e+04 | 0.32 |
| 2017-03-09 04:16:37 | 1.03 | 40.89 | 2.7e-06 | 2.3e-05 | 0.00012 | 8.46e+06 | 979 | 8.64e+03 |
| 2022-07-28 01:36:07 | 1.01 | 8.024 | 2.5e-07 | 2.1e-06 | 1.1e-05 | 9.63e+07 | 1.02e+03 | 9.42e+04 |
| 2009-04-10 18:42:45 | 1.37 | 1.939 | 5.7e-08 | 4.4e-07 | 2e-06 | 2.07e+08 | 412 | 5.01e+05 |
| 2021-05-06 05:54:26 | 0.69 | 1.53 | 1.1e-07 | 1.1e-06 | 6.3e-06 | 5.05e+08 | 3.19e+03 | 1.58e+05 |
| 2015-02-17 13:19:50 | 0.61 | 1.249 | 2.4e-07 | 2.5e-06 | 1.5e-05 | 3.21e+08 | 4.72e+03 | 6.8e+04 |
| 2022-07-22 00:16:18 | 0.96 | 0.01648 | 5.7e-10 | 5e-09 | 2.5e-08 | 4.69e+10 | 1.19e+03 | 3.94e+07 |
| 2026-01-30 10:25:37 | 0.32 | 3421 | 0.0035 | 0.046 | 0.26 | 9.61e+04 | 3.34e+04 | 2.88 |
| 2015-03-11 06:18:59 | 0.92 | 0.01782 | 1.3e-09 | 1.2e-08 | 6.1e-08 | 2.25e+10 | 1.38e+03 | 1.63e+07 |
| 2013-07-31 03:50:13 | 0.98 | 0.007395 | 4.7e-10 | 4.1e-09 | 2.1e-08 | 5.42e+10 | 1.13e+03 | 4.8e+07 |
| 2025-06-26 00:37:20 | 1.11 | 0.01256 | 3.1e-10 | 2.6e-09 | 1.3e-08 | 6.15e+10 | 786 | 7.82e+07 |

Table 9. Exploratory sensitivity to origin assumptions. The table compares required number densities with model-dependent speed-only posterior quantities to show the astrophysical scale of the assumptions. These are not full-state posterior probabilities because radiant uncertainty and, for Polar-IM, the independent stereo state are not modelled jointly.

Within this calculation, Polar-IM is the only event whose required metre-scale density overlaps the steepest published size-distribution extrapolation. The 2026 January 30 and 2014 January 8 events are the next closest, and the remaining candidates require still larger densities. This ordering is useful for follow-up priority. It does not establish an interstellar origin for any event.

A hyperbolic trajectory also has a third possible origin. Peña-Asensio et al. (2024b) show that Oort-cloud perturbations can send Solar System objects onto hyperbolic Earth-impact trajectories. Such objects should have small positive excess speeds rather than the tens of kilometres per second expected for the most extreme interstellar cases. In Table 9, this channel is most relevant to the six events with heliocentric excess speeds below 10 km $s^{-1}$. It is not a plausible explanation for Polar-IM at 31.6 km $s^{-1}$ or the 2014 January 8 event at 44.9 km $s^{-1}$. We do not quantify the Oort-cloud flux here, but it is an astrophysical alternative to both measurement error and interstellar origin for the weaker candidates.

The size-distribution calculation is uncertain for several reasons. The steepest slope is constrained by meteor-survey limits, the kilometre-scale anchor is based on very few detections, impact-energy changes affect the inferred radius, and the event-level likelihood is incomplete. We therefore use this calculation as a population-scale consistency check and a way to compare follow-up priorities.

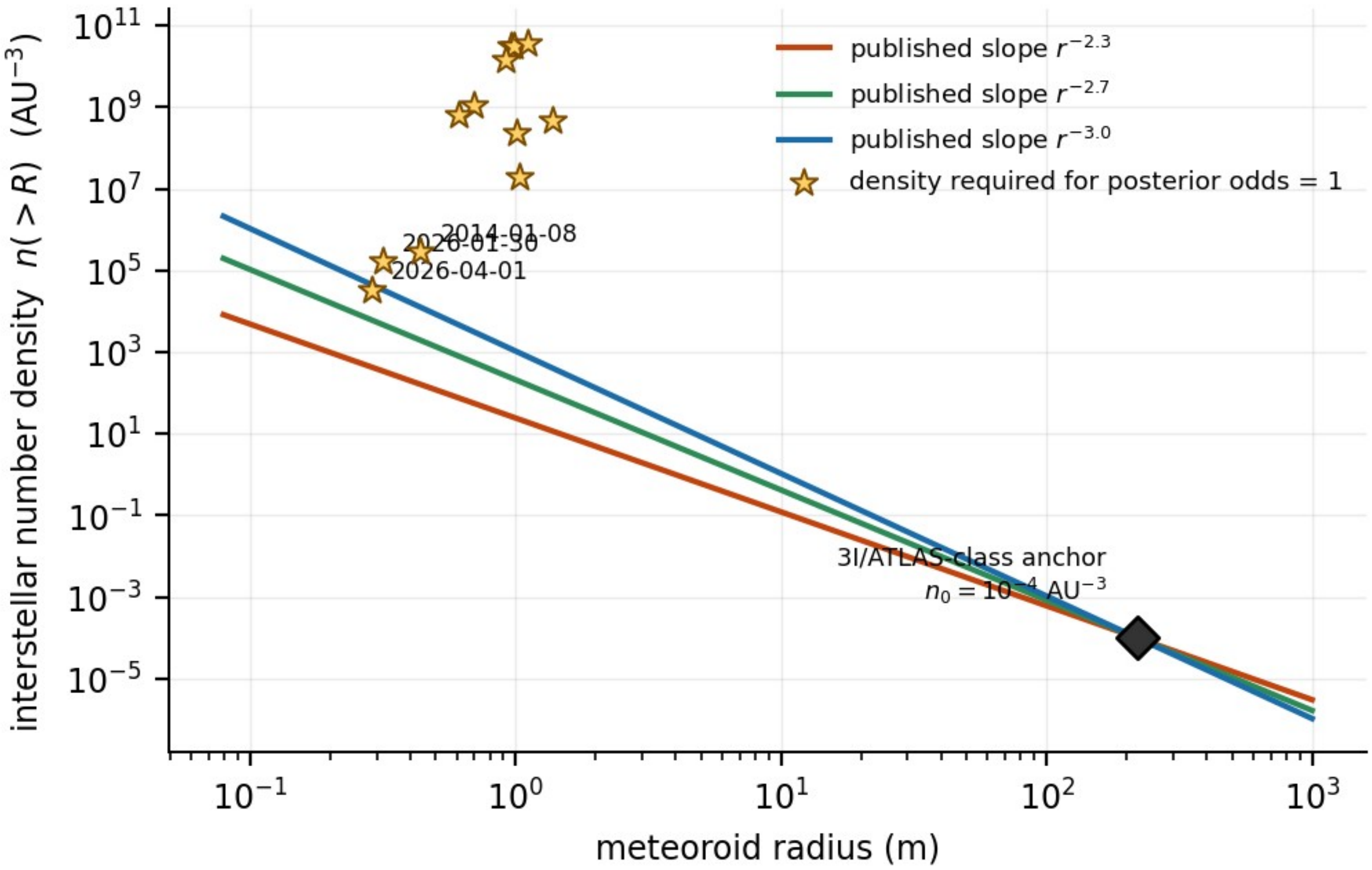

Figure 9. Published interstellar number-density extrapolations compared with the density required by the exploratory speed-only model. This is a prior-predictive stress test, not a determination of origin.

### 5.5 Infrasound source-location check

Section 4.9 shows why velocity direction is essential to the orbital classification. The infrasound data of Silber et al. (2025) do not measure the velocity radiant, but they do provide an independent angular check on the reported source location through array back azimuth. We use the dataset for that purpose and for context on large solution discrepancies.

The Silber et al. (2025) event-level database contains 362 infrasound detections of 137 CNEOS bolides between 2000 July and 2023 May, recorded at International Monitoring System and regional arrays. Each detection includes the observed array back azimuth and the theoretical back azimuth from the CNEOS event position. The release also includes signal period, amplitude, celerity, station distance, and event-level estimates of impactor density, diameter, mass, dynamic strength, Tisserand parameter, entry angle, and luminous begin and end altitudes. The back azimuth is measured independently of CNEOS, so it provides a large-sample check on source geolocation.

Figure 10 summarizes the comparison. Of the 362 detections, 331 have both observed and theoretical bearings. The observed-minus-theoretical residual has a median of -0.20°, a robust scatter of 4.15°, and a median absolute value of 2.83°. The tail is substantial: 42 detections (12.7%) exceed 10°, 22 (6.6%) exceed 15°, and the largest residual is 48.6°. This residual contains both any CNEOS position error and acoustic propagation away from the great-circle path. It therefore places an upper bound on the CNEOS contribution rather than measuring it directly.

We also reduce the data to one median residual per event so that a bolide detected by several arrays is not counted several times. This leaves 121 events, with eight above 10°. The Jeffreys mean is 6.97%, with a 95% interval of 3.18-12.09%. The rate is similar to the GLM position-outlier rate, but it is not a CNEOS radiant-error rate: the observable is an acoustic source bearing and includes atmospheric propagation. The pooled cross-modal value (16/255, 6.45%) is therefore retained only as a sensitivity calculation.

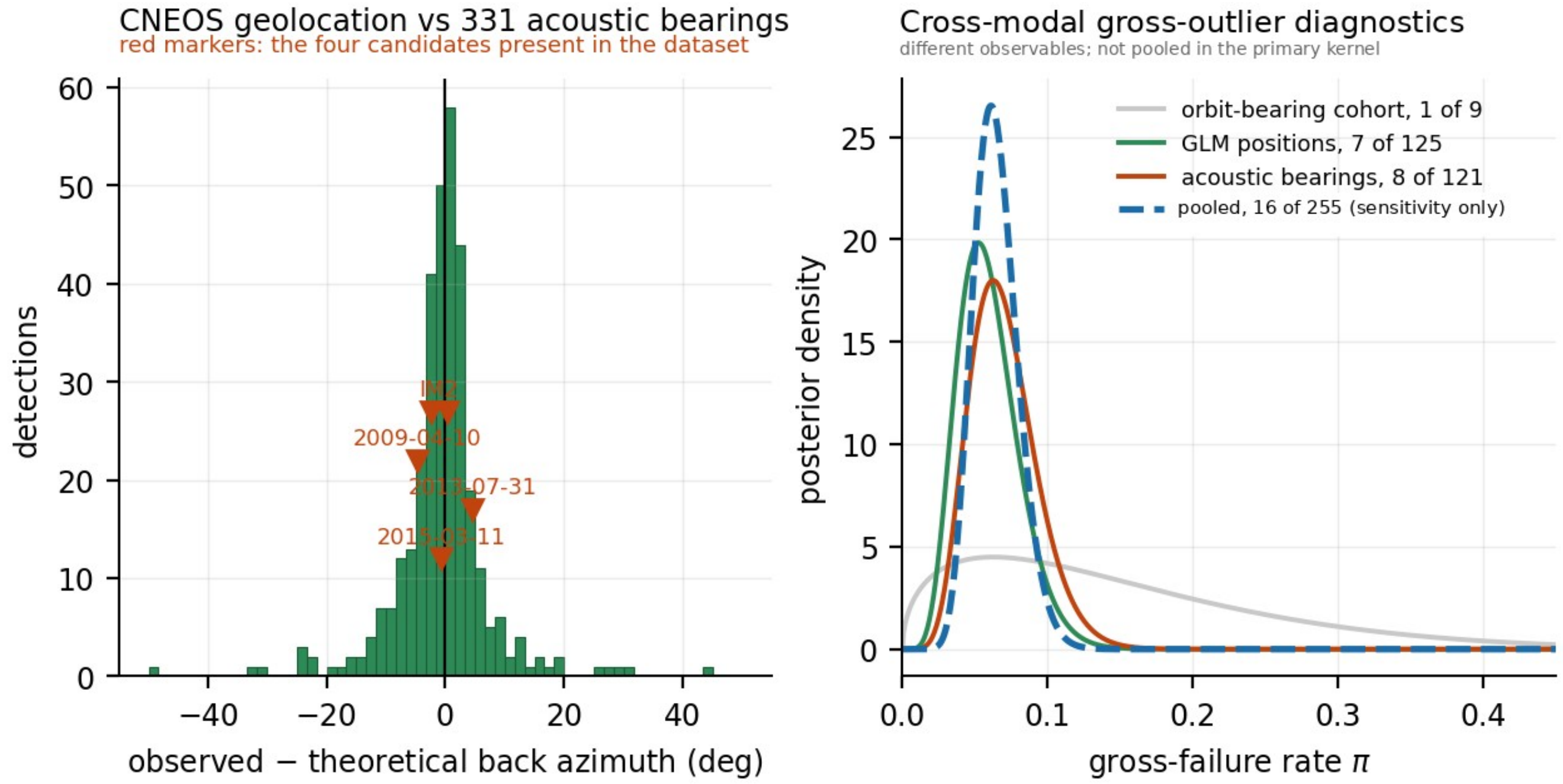

Figure 10. Left: observed-minus-CNEOS-derived infrasound back-azimuth residuals for 331 detections. Right: separate posteriors for large discrepancies in the orbit-bearing radiant, GLM position, and acoustic-bearing channels. Their similarity is shown only as a cross-modal comparison; the primary analysis does not assume a common failure probability.

Four nominal or stage-sensitive candidates appear in the acoustic database. IM1 was detected by two arrays from substantially different viewing azimuths, and both bearings agree with the reported CNEOS source location. This makes a gross location error less likely for that event, but it does not validate the direction of the CNEOS velocity vector.

| Event (UTC) | Rank | Detections | Station range (km) | Observed − theoretical bearing (°) | Required radiant error (°) |
|---|---|---|---|---|---|
| 2014-01-08 17:05:33 | 1 | 2 | 1750, 2525 | +0.35, −2.48 | 43.7 |
| 2009-04-10 18:42:45 | 5 | 1 | 1788 | −4.82 | 34.6 |
| 2015-03-11 06:18:59 | 10 | 1 | 1703 | −0.76 | 2.3 |
| 2013-07-31 03:50:13 | 11 | 1 | 1511 | +4.60 | 1.6 |

Table 10. Candidate events with independent acoustic bearings. Back azimuth checks the source location rather than the velocity direction, so the table constrains gross geolocation failure without directly adjudicating hyperbolic provenance.

A back azimuth points from the array to the acoustic source; it does not measure the meteor's velocity radiant. For IM1, agreement between two arrays argues against a large source-location error, but it cannot rule out a velocity-direction error. For the marginal 2013 and 2015 events, the radiant change needed to alter the orbital classification is smaller than the practical angular precision of the acoustic test. We therefore use infrasound as a location check, not an orbit measurement.

The same infrasound release also shows how strongly the exploratory size-distribution calculation depends on assumed physical properties. For IM1, using the released density/radius choice changes the required-to-available interstellar density ratio from 9.36 to 1.53. That change is not evidence for origin. The density is model-derived, the size-distribution extrapolation is uncertain, and the acoustic bearing constrains location rather than velocity direction. It simply shows that the provenance calculation can be as sensitive to physical assumptions as to the statistical tail model.

The release gives luminous begin and end altitudes for 86 events. Their medians are 40.3 and 25.6 km, with a median path span of 13.8 km. These measurements are the kind of input needed for the altitude-resolved stage model described in Section 4.3. We note their availability but do not build that model here.

### 5.6 What drives the infrasound residual?

The paired bearings also let us ask what drives the acoustic residuals. Figure 11 tests three possibilities: reported speed, station distance, and entry geometry.

First, there is no detectable relation with reported speed. CNEOS gives a speed for 271 acoustic detections, ranging from 9.8 to 44.8 km $s^{-1}$, and the absolute back-azimuth residual has Spearman $\rho$ = -0.039 with p = 0.52. Source-geolocation agreement does not obviously worsen across this range. This does not test the velocity radiant, and the high-speed gap remains: only eight detections exceed 30 km $s^{-1}$ and none approaches the 69 km $s^{-1}$ of Polar-IM. The direct calibration still ends at 38.6 km $s^{-1}$.

Second, station distance matters because acoustic propagation contributes to the residual. In a structured, windy atmosphere, sound can be refracted away from the great-circle path (Silber 2024). An illustrative fit to five equal-count distance bins, $\sigma(D)^2 = \sigma_0^2 + (aD)^2$, gives $\sigma_0 = 3.45°$ and a = 0.56° per 1000 km. At the median range of 2820 km, the propagation-scale contribution is 1.58°. Residuals above 10° also become more common at the largest distances. The fitted distance-independent term is therefore only an upper bound on distance-independent geolocation/systematic scatter, not a clean CNEOS error term.

Third, we test entry geometry. Bolide entry angle affects infrasound radiation and detectability, and steeper entries are preferentially detected (Ronac Giannone & Silber 2026). Apparent arrival direction has also been studied directly (Silber 2025). Among the 271 detections with an entry angle, the absolute residual is uncorrelated with it ($\rho = -0.069$, p = 0.26). We therefore find no evidence that entry angle is driving the bearing residual in this sample.

Finally, 86 events include luminous begin and end altitudes as well as the reported CNEOS altitude. The reported altitude lies at a median 44.5% of the way down the luminous path, with an interquartile range of 30-58%. The median begin, reported, and end altitudes are 40.3, 33.3, and 25.6 km. This supports a role for measurement stage, but it does not identify the exact epoch represented by the velocity components. CNEOS calls them entry velocities while also describing returned fields as relative to peak brightness. An event-specific stage model is therefore preferable to a universal +0.9935 km $s^{-1}$ correction.

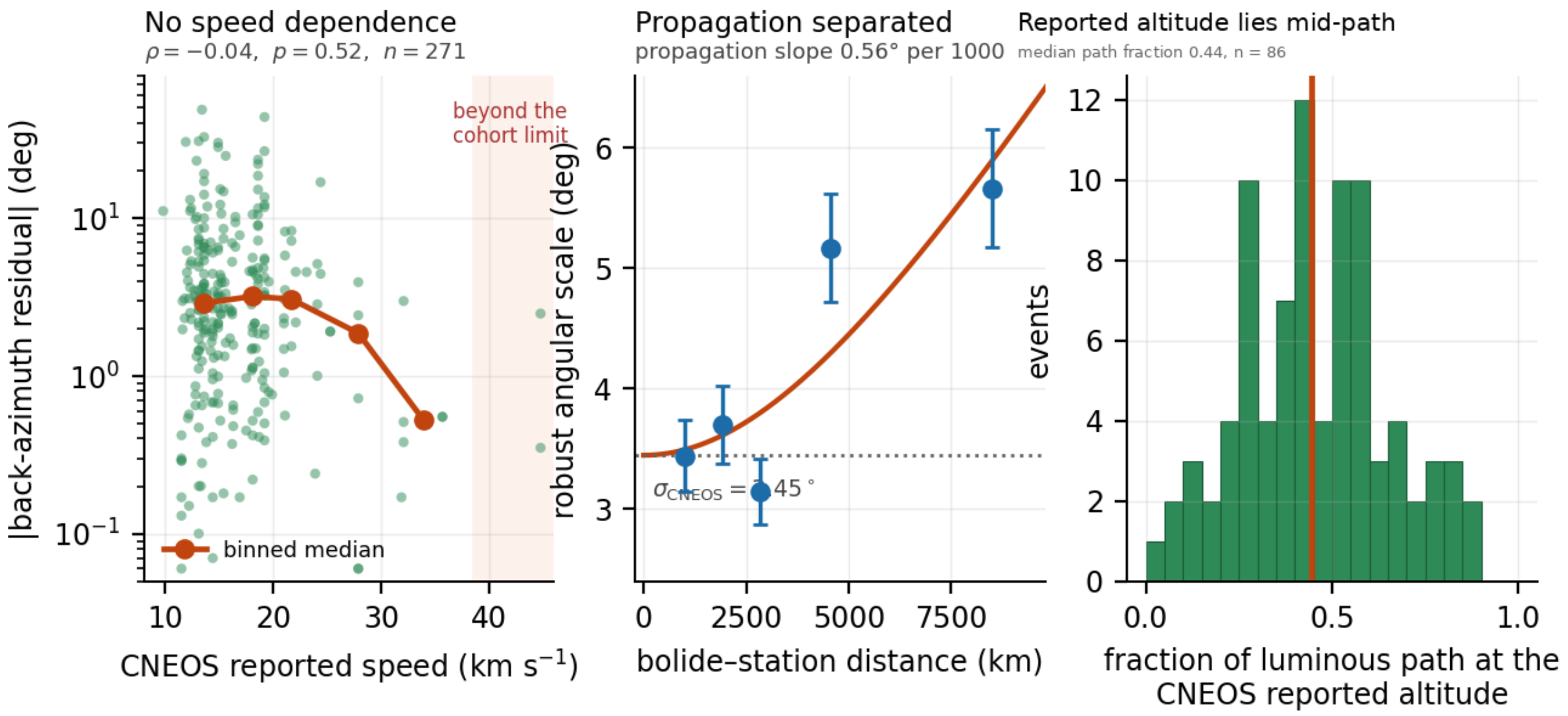

Figure 11. Structure of the acoustic-bearing residuals. Left: absolute back-azimuth residual versus reported speed, with no detectable trend over the sampled range. Centre: robust angular scale versus station distance, with an illustrative distance-independent plus propagation-scale fit. Right: the reported CNEOS altitude within the luminous path for 86 events. The right panel describes altitude geometry; it does not determine the epoch represented by the published velocity vector.

## 6. How This Calibration Differs from Earlier Work

### 6.1 Interpreting published calibration widths

Our calibration differs from the kernel used in recent candidate studies mainly in how the published summary statistics are interpreted. The source data are the same. Table 4 of Peña-Asensio et al. (2025) is captioned “Median and (upper and lower) 1σ standard deviations of the DD values and orbital element, velocity, and geocentric radiant errors for the 18 calibrated fireballs from CNEOS”. For the low-discrepancy group, Table 11 reproduces three values: a median velocity error of 0.55 km s⁻¹ with +0.37/−0.45, a median right-ascension error of 1.35° with +1.12/−0.79, and a median declination error of 0.84° with +1.00/−0.79.

| Quantity | PA25 low-discrepancy median | 1σ spread | Use in Cloete & Loeb (2026a,b) | Implied σ if the errors are half-normal | This work, measured |
|---|---|---|---|---|---|
| velocity error | 0.55 km s⁻¹ | +0.37 / −0.45 | Gaussian σ | 0.82 km s⁻¹ | 1.29 km s⁻¹ (RMS) |
| right-ascension error | 1.35° | +1.12 / −0.79 | Gaussian σ | 2.00° | — |
| declination error | 0.84° | +1.00 / −0.79 | Gaussian σ | 1.25° | 0.74° per axis (core) |

Table 11. Published low-discrepancy error summaries from Peña-Asensio et al. (2025), the way they have been used in recent candidate analyses, and an illustrative conversion if a median error magnitude were treated as a half-normal quantity. The half-normal column divides the median by 0.6745; this is an illustration, not a refit.

These published values are medians of positive per-event error magnitudes with asymmetric spreads. They are not a fitted zero-mean Gaussian standard deviation. Using the median magnitude as σ in $N(0, \sigma^2)$ is therefore an additional modelling step. For example, a half-normal distribution with a median of 0.55 km s⁻¹ would correspond to a parent Gaussian standard deviation of 0.82 km s⁻¹. That is about 64% of the 1.29 km s⁻¹ modern high-quality RMS measured here. We are not adopting a half-normal model; this calculation simply shows why 0.55 km s⁻¹ is not automatically a Gaussian standard deviation.

The asymmetric spreads point in the same direction: one symmetric scale around zero does not describe the published error magnitudes. This motivates testing non-Gaussian tails. It does not, by itself, determine the two-component mixture used here; that choice is motivated by the 16.3° radiant discrepancy in the modern direct calibration.

This is a difference in statistical interpretation, not a criticism of Peña-Asensio et al. Their summary statistics are reported transparently, and using a central published value as an error-kernel width is a practical approximation when the full event-level likelihood is unavailable. Peña-Asensio et al. also caution against over-interpreting hyperbolic CNEOS candidates. They note that CNEOS reports few fast fireballs and find significant inaccuracies in more than 22% of CNEOS events in a random sample. Our aim is to show how alternative interpretations of the kernel change the candidate evidence.

### 6.2 Large discrepancies in other CNEOS observables

Several studies have found substantial differences between some CNEOS solutions and independent measurements, especially in radiant or orbital quantities. Our modern direct sample contains one 16.3° radiant discrepancy in nine comparisons. GLM positions and infrasound bearings also show large discrepancies at several-percent rates, although they measure different observables. Together, these results justify testing a robust error tail. They do not support a single universal failure probability for every CNEOS field.

Hajduková et al. (2024) found no population-level evidence for interstellar fireballs in CNEOS, and our population checks are compatible with that result. The stage-sensitivity count is not unusual under the conservative null, and published interstellar size distributions predict far fewer than one genuine event in the modern vector-bearing sample. That does not make the event-level follow-up list meaningless. Several states have large model-free error margins and are worth independent measurement. Population-level scarcity and event-level follow-up therefore address different questions.

The same paper provides a useful independent check on Section 4.11. Its authors note that, during publication, CNEOS changed the timestamps of half of the hyperbolic events by one second and revised energies. Our archived-snapshot comparison finds the same behaviour. The median re-timing in our comparison is 1.0 second, exactly matching that report.

### 6.3 Relation to recent interstellar-meteor candidate studies

Relative to Peña-Asensio et al. (2025), we make several methodological changes. We confirm the post-2018 improvement in CNEOS fidelity, separate reference provenance from observing epoch, treat speed and radiant errors separately, and propagate uncertainty in the directly observed large-direction-error rate. We also find a mean speed offset compatible with a difference in atmospheric measurement stage, although a velocity-definition difference cannot be excluded (Appendix A.4). Rather than correcting every event, we show that offset as a sensitivity test.

For Cloete & Loeb (2026a, 2026b), we reproduce CNEOS-22 and Polar-IM from the current catalogue and recover the published CNEOS-only reductions under their assumptions. Silber et al. (2026) provide an independent same-event measurement for Polar-IM. Their directions are similar to CNEOS, but the speeds are about 18% lower and lie close to the parabolic threshold for those directions. In a simple one-dimensional Gaussian screen, roughly one-third of the stereo probability is unbound. A joint CNEOS-plus-stereo posterior would require the stereo covariance and a validated likelihood for the large CNEOS discrepancy, which are not available. The practical conclusion is that Polar-IM deserves a full uncertainty-bearing stereo reconstruction before its origin is decided.

Barghini et al. (2025) make a related methodological point from ground-based catalogues: formal eccentricity errors near the parabolic boundary can be unreliable. That caution matters even more for a catalogue that publishes no per-event errors. It is one reason we report the model-free required-error statistic alongside probability-based sensitivity calculations.

## 7. Supporting Public-Data Checks

We also examined other public meteor, fireball, radar, satellite, and infrasound datasets. These sources have a supporting role. Most do not measure the same quantity as the CNEOS velocity vector, so we do not combine them into one exposure or use them to enlarge the direct calibration sample. The checks below are included because they test specific questions about quality selection, geolocation, or data-product limitations.

### 7.1 Quality cuts and nominal hyperbolic fractions

Quality cuts strongly affect nominal hyperbolic fractions in the large catalogues. In GMN (Vida et al. 2020, 2021), 280,201 of 3,415,540 trajectories are nominally hyperbolic (8.20%); after the study quality screen, the fraction is 3.66%. In SonotaCo (SonotaCo et al. 2021), it falls from 11.06% to 4.98%. In PANSY (Vierinen 2026), it falls from 4.24% to 0.65% across explicit Level 3 selections. In FRIPON (Colas et al. 2020), it falls from 10.12% to 1.82%, and no FRIPON solution is more than three reported eccentricity standard deviations above unity. CAMSv3 falls from 11.88% to 7.12%. SAAMER gives 10.73% and AMOR 10.00%, although neither release provides per-record errors for a formal significance test. In the speed-matched range $10 \leq vg < 30$ km s$^{-1}$, the CNEOS nominal rate is 4/240 (1.667%, Wilson 0.650-4.206%), compared with 0.336% for GMN and 0.105% for quality-screened PANSY. These catalogues use different instruments, populations, and reduction pipelines. The comparison shows the effect of quality selection; it is not a direct estimate of the CNEOS false-positive rate.

### 7.2 Common-event location checks

The GLM bolide catalogue (Smith et al. 2021; Ozerov et al. 2024) gives 130 one-to-one CNEOS matches within 60 s and 300 km. Of these, 69 have a public CNEOS velocity vector. Twenty-three catalogue-level matched positions are exactly equal to rounded CNEOS coordinates, so they cannot be assumed independent. Using the native satellite attachment positions gives a median event-level horizontal separation of 39.46 km. This is a source-location check only; it does not calibrate the CNEOS velocity vector.

### 7.3 Public-data limitations

Some public products are not suitable for the uncertainty calculation we would like to perform. The distributed CAMO/EMCCD covariance arrays (Western Meteor Physics Group 2026), as parsed here, fail the stated numerical admissibility checks for nearly all rows, so we do not propagate them. The four-pipeline Geminid study of Shober et al. (2026) shows that formal velocity uncertainties for the same events can differ substantially between reduction pipelines even when trajectory RMS values are similar. MAARSY/PANSY timing fields (Vierinen 2025, 2026) also need source-specific interpretation. These points concern the released products as ingested here, not the quality of the underlying observations or unpublished processing.

### 7.4 Held-out event checks

Several held-out events provide useful spot checks. For the 14 August 2026 Washington event, the reported starting speed is 14.6 km s$^{-1}$ and the CNEOS component norm is 22.899 km s$^{-1}$; the beginning altitude is 77.9 km compared with a CNEOS brightness altitude of 30 km. The sign of the speed difference is opposite to ordinary deceleration. For the 17 March 2026 Ohio event, the corresponding speeds are 17.5 and 14.915 km s$^{-1}$, consistent with slowing. The 2023 May 20 Queensland event is the nineteenth calibrator. Its speed residual is −1.579 km s$^{-1}$ and its radiant separation is 1.295°, both within the modern calibrated core. It is the strongest single piece of evidence in this set that the modern CNEOS vector is accurate at the level claimed here. Florida/Grand Bahama (Hughes et al. 2022) gives a twentieth speed-channel comparison at −3.351 km s$^{-1}$, but its tabulated radiant equinox is not explicit, so it is not included in the frame-harmonised cohort.

## 8. Discussion

The modern CNEOS error core is more constrained than the rare-error tail, but remains uncertain with only nine events. Nine modern high-quality events define a narrow speed and radiant core, but the 16.3° discrepancy for 2019 MO shows that large direction errors do occur. GLM geolocation and infrasound bearings show several-percent large discrepancies in other CNEOS-related observables. Those external datasets support sensitivity tests with a large-error tail model. The direct orbit-bearing sample remains the relevant calibration for velocity direction. Appendix A shows that this distinction controls the event-level results: the core parameters are robust to the reference-error assumptions, whereas P(bound) for the strongest candidates depends almost entirely on the treatment of the single 16.3° event and, for Polar-IM, on how speed errors scale with speed.

The mean CNEOS-minus-reference speed offset is approximately -1 km $s^{-1}$, close to the EFN difference between pre-atmospheric and luminous-stage speeds. A measurement-stage contribution is therefore plausible, but it is not demonstrated: an EFN-calibrated drag model matches the sign and order of magnitude of the mean offset but not its event-to-event variation, so a velocity-definition difference remains possible. A single global shift should not be applied automatically to every event because the CNEOS field description is ambiguous, EFN deceleration varies widely, and the fastest CNEOS events are reported high in the atmosphere. The primary result is the uncorrected seven-event set; the five extra crossings produced by +0.9935 km $s^{-1}$ show the sensitivity to stage.

Polar-IM is the clearest same-event test. CNEOS and the GLM-LI stereo reconstruction agree fairly well in direction but disagree strongly in speed. The stereo speeds lie near the direction-specific parabolic thresholds. If the quoted spreads are treated as Gaussian scales in a simple one-dimensional calculation, approximately one-third of the stereo probability is unbound. The missing full stereo covariance means that the direct disagreement between the measurements is more informative than a highly precise CNEOS-only tail probability.

The next step is more high-speed common events with independent three-dimensional trajectories and quantified uncertainties. The 2026 April 1 and 2026 January 30 events are especially useful because both lie above the 38.6 km $s^{-1}$ limit of the direct calibration sample. Infrasound can help with source location and energetics, but a back azimuth is not an independent measurement of the velocity radiant.

Three follow-up tasks would most improve the evidence. The 2026 April 1 GLM-LI observations need a covariance-aware reconstruction with an explicit deceleration model. A contemporaneous CNEOS-25 payload should be recovered if possible. The 2026 January 30 New Zealand event also needs an independent trajectory. More generally, high-speed CNEOS events should be targeted for independent trajectories, and per-event CNEOS uncertainty or quality metadata would remove much of the present instrumental ambiguity.

The most direct way to resolve these questions is not statistical. The limitations identified here include the poorly measured tail, the extrapolation above 40 km $s^{-1}$, the ambiguity between deceleration and velocity definition, and the revision of a published vector discussed in Section 4.6, and are aggravated by the absence of per-event uncertainty information in the public CNEOS release. The catalogue is derived from U.S. Government sensors whose data are made available by the U.S. Space Command and Department of Defense (Brown et al. 2002), and their internal processing may carry a measure of solution quality that is not released. For the 2014 January 8 event, the U.S. Space Command has stated in an official letter that "the velocity estimate reported to NASA is sufficiently accurate to indicate an interstellar trajectory" (Siraj & Loeb 2022a), without publishing the uncertainty on which that judgement rests. Uncertainty information therefore exists and is assessed internally; what is missing is its release. We therefore recommend that the agencies responsible for these sensors make public, at least for events proposed as interstellar candidates, the full velocity covariance or component uncertainties and correlations, the velocity epoch and reference frame, available systematic-error and quality information, together with a record of any later revisions to the published values. Quality flags would help stratify calibration samples; reliable event-level inference also needs quantified errors and validation of rare failures.

Broader statements about impact energy or flux require an event-resolved measurement-stage model. The present data show that measurement stage can matter. They do not support shifting every CNEOS speed by the same +1 km $s^{-1}$ amount.

## 9. Conclusions

The main conclusion is simple: CNEOS is useful for finding unusual events, but the catalogue alone does not securely identify an interstellar meteor. The evidence has distinct levels. The same-event residuals and the count of nominally hyperbolic states are empirical results for the selected sample and frozen catalogue. The decomposition of measurement variance and the calculated bound fractions depend on the adopted reference-error and perturbation models. The physical frequency of extreme errors, their behaviour beyond the calibrated speed range, and the interstellar origin of individual candidates remain insufficiently constrained by these data.

In the modern high-quality calibration subset (n = 9), the speed residual has a mean of -1.00 km $s^{-1}$ and a standard deviation of 0.86 km $s^{-1}$. The variance-component estimate for the CNEOS contribution is 0.77 km $s^{-1}$. The radiant residuals have a 0.74° per-axis core and one 16.3° large discrepancy. The core error scale is therefore better constrained than the extreme direction-error rate, but its sampling uncertainty remains substantial. The CNEOS speed scale is robust to the assumed reference-trajectory errors (0.67–0.88 km $s^{-1}$ across the alternatives tested), but the tail is not measured well enough to fix event-level probabilities for the strongest candidates.

The mean speed offset is consistent with the EFN difference between pre-atmospheric and luminous-stage speeds, but an EFN-calibrated drag model does not reproduce its event-to-event variation, so a velocity-definition difference cannot be excluded. The available metadata do not support a universal +0.9935 km $s^{-1}$ correction. With the published vectors, seven states are nominally hyperbolic. Five more cross the boundary only in the atmospheric-stage sensitivity test.

Polar-IM is the strongest post-2018 CNEOS anomaly. The independent GLM-LI stereo reductions give 57.3 ± 2.0 and 56.7 ± 3.0 km $s^{-1}$, close to the parabolic thresholds for their measured directions. The independent measurement therefore leaves the event unresolved rather than confirming the CNEOS hyperbolic solution. Its CNEOS-only probability is also conditional on the assumption that absolute speed errors measured below 40 km $s^{-1}$ apply at 69 km $s^{-1}$; under a fractional error law, its required speed change is comparable to the largest fractional error seen in the modern calibration.

GLM geolocation and infrasound back azimuth show several-percent large discrepancies in their own observables, but neither directly measures the CNEOS velocity-radiant tail. The archived 2019 snapshot also shows that some CNEOS fields change over time. Candidate studies should therefore preserve the catalogue payload and retrieval metadata used in the analysis.

Taken together, the results support a cautious but useful role for CNEOS. The catalogue can rank unusual events and identify good follow-up targets. It cannot, from one velocity vector alone, distinguish a genuinely unbound metre-scale object from a catalogue error. For Polar-IM, the decisive test is a joint three-dimensional reconstruction of the independent stereo observations with propagated uncertainty. Release of per-event CNEOS velocity uncertainties, at least for proposed interstellar events, would substantially improve these assessments, while calibration of systematic errors and independent trajectories would still be needed.

### Acknowledgments

This work was supported in part by the Galileo Project at Harvard University. The study uses the NASA/JPL CNEOS Fireball API, GOES Geostationary Lightning Mapper (GLM), Meteosat Third Generation Lightning Imager (LI), and the public infrasound-array products described in the cited releases.

During manuscript preparation, OpenAI’s ChatGPT and Codex, and Anthropic’s Claude were used for language editing, organization, code drafting, computational cross-checking, and consistency checks. Source provenance, intermediate results, and relevant computational outputs were retained for verification.

## Data Availability

The source datasets used in this study are publicly accessible and are identified in the manuscript and cited releases. The analysis separates the seven-event nominal set from the atmospheric-stage sensitivity cases, keeps direct radiant calibration distinct from cross-modal checks, and treats the CNEOS-only and Polar-IM stereo calculations separately. Random seeds, required-error statistics, catalogue-version checks, and exploratory prior-sensitivity calculations were retained for reproducibility. The scripts used for the sensitivity analyses in Appendix A, together with their input tables, are provided as supplementary material. The supplement reproduces the sensitivity analyses of Appendix A, Table 4 and Figure A1 from the included input tables and lists the tested software environment; the orbital reductions of Section 3.1 enter it as reduced tables.

# Appendix A. Sensitivity analyses

This appendix collects the sensitivity tests summarised in Sections 4.1–4.3 and 4.8. All calculations use the uncorrected CNEOS vectors and are carried out in the geocentric-asymptote frame in which the calibration residuals are measured. For each event, Earth's heliocentric speed is approximated by vis-viva using the local solar escape speed and a fixed Earth semimajor axis, and the angle between the geocentric asymptote and Earth's velocity is recovered from the reduced heliocentric speed. This reproduces the margins of Section 4.9 (for Polar-IM, 23.0° of radiant rotation and 10.6 km $s^{-1}$ in reported speed, equivalent to 10.8 km $s^{-1}$ in vg). Speed errors are zero-mean Gaussian deviates unless stated otherwise; this deliberately leaves the fitted negative bias unapplied to the nominal vectors. It is a sensitivity convention rather than a fully fitted generative model. Direction errors are rotations of Rayleigh-distributed magnitude and uniform azimuth. P(bound) is the fraction of $4\times10^6$ draws with $v_\odot < v_{esc}$; $< 10^{-6}$ denotes no bound draw, and a single bound draw corresponds to $2.5\times10^{-7}$. With zero bound draws, the exact one-sided 95% binomial upper limit is $7.49\times10^{-7}$ for $4\times10^6$ independent trials; the notation $<10^{-6}$ is a rounded Monte Carlo limit under that kernel, not an empirical bound on the physical error tail. The same speed width is used in both angular components, and speed and direction are conditionally independent. Their joint tail dependence is not calibrated.

### A.1 $\sigma_{CNEOS}$ and the reference-error assumptions

The variance-components model of Section 3.2 is refitted under the reference-error scenarios in Table A1. For the adopted joint fit, each profile point re-optimises both the common bias and the network-reference standard deviation, including its zero boundary. In the fixed-reference, inflated-astrometric and correlation scenarios, the listed reference scales are held fixed and only the bias is profiled. The leave-one-out fits re-estimate the bias and network scale each time. Likelihood intervals use the one-parameter chi-square cutoff and are approximate with n = 9 and a variance component that can reach zero. The class-A-only row instead reports a sample standard deviation with its normal-theory chi-square interval.

| Assumption | $\sigma_{ref,A}$ (km $s^{-1}$) | $\sigma_{ref,N}$ (km $s^{-1}$) | ρ | b (km $s^{-1}$) | $\sigma_{CNEOS}$ (km $s^{-1}$) | 68% interval | 95% interval |
|---|---:|---:|---:|---:|---:|---:|---:|
| Network error fitted (adopted) | 0.02 | 0.33 | 0 | -1.03 | 0.77 | 0.54–1.04 | 0.42–1.39 |
| Network error fixed | 0.02 | 0.00 | 0 | -1.00 | 0.81 | 0.65–1.04 | 0.54–1.39 |
| Network error fixed | 0.02 | 0.10 | 0 | -1.00 | 0.80 | 0.65–1.04 | 0.54–1.39 |
| Network error fixed | 0.02 | 0.20 | 0 | -1.01 | 0.79 | 0.63–1.03 | 0.52–1.38 |
| Network error fixed | 0.02 | 0.50 | 0 | -1.06 | 0.73 | 0.56–0.98 | 0.45–1.35 |
| Network error fixed | 0.02 | 0.75 | 0 | -1.11 | 0.69 | 0.52–0.94 | 0.41–1.32 |
| Network error fixed | 0.02 | 1.00 | 0 | -1.16 | 0.67 | 0.50–0.93 | 0.40–1.33 |
| Astrometric error inflated | 0.10 | 0.33 | 0 | -1.02 | 0.77 | 0.60–1.01 | 0.48–1.37 |
| Astrometric error inflated | 0.20 | 0.33 | 0 | -1.02 | 0.76 | 0.59–1.00 | 0.46–1.36 |
| Correlated errors | 0.02 | 0.33 | -0.5 | -1.07 | 0.70 | 0.54–0.94 | 0.44–1.29 |
| Correlated errors | 0.02 | 0.33 | +0.5 | -0.97 | 0.88 | 0.71–1.12 | 0.59–1.48 |
| Class A only (n = 4) | — | — | — | -1.33 | 0.82 | — | 0.46–3.05 |
| Leave-one-out (range) | 0.02 | refitted | 0 | — | 0.45–0.85 | — | — |

Table A1. Sensitivity of σCNEOS to the reference-error assumptions for the nine modern high-quality events. σref,A and σref,N are the astrometric and network reference errors. Across all reference-error alternatives σCNEOS stays between 0.67 and 0.88 km $s^{-1}$, inside the 95% statistical interval of the adopted fit. The class-A row uses the four astrometric events only and has negligible reference error on the displayed scale. In the leave-one-out row the lowest value (0.45 km $s^{-1}$) is obtained when 2018 LA is removed; both nuisance parameters are re-estimated.

## A.2 Core, tail and the 16.3° event

Table A2 gives P(bound) for the seven nominal events under three interpretations of 2019 MO. S1 excludes it from the angular fit and uses the 0.74° per-axis core alone. S2 treats it as a draw from the same population and fits one Rayleigh law to all nine events (3.91° per axis). S3 treats it as a member of a second population: a mixture of the core and a gross component of width σgross = 11.53° per axis (the maximum-likelihood value for the 16.31° separation), with weight π drawn from the Jeffreys posterior Beta(1.5, 8.5). Further rows vary σgross, fix π at the 6.45% cross-modal value, and replace the Gaussian speed error with a Student-t law (ν = 3) of the same standard deviation. The speed error is 0.77 km $s^{-1}$ throughout. The last row gives the fraction of the S3 probability contributed by the gross component, computed as 1 − (1 − E[π])PS1/PS3, where E[π] = 0.15. The speed fit is held fixed across these three angular scenarios; removing 2019 MO from the joint speed fit instead gives σCNEOS = 0.63 km $s^{-1}$ (Table A1 supplementary leave-one-out output). Drawing a new π for each Monte Carlo realization marginalises over π and yields a mean, not a credible interval for P(bound). Holding component shapes fixed, varying the shared π over its central 95% posterior interval gives P(bound) ranges of approximately $1.8\times10^{-6}$–$6.0\times10^{-5}$ for IM1, $2.1\times10^{-4}$–$7.1\times10^{-3}$ for Polar-IM, $6.3\times10^{-4}$–$2.1\times10^{-2}$ for IM2, and $9.7\times10^{-4}$–$3.3\times10^{-2}$ for CNEOS-22. These intervals include uncertainty in π only; uncertainty in component widths, bias, epoch transfer and Monte Carlo estimates is additional.

| **Scenario** | **2014-01-08 IM1** | **2026-04-01 Polar-IM** | **2017-03-09 IM2** | **2022-07-28 CNEOS-22** | **2009-04-10** | **2021-05-06** | **2015-02-17** |
|---|---|---|---|---|---|---|---|
| S1 outlier excluded (core only) | $<10^{-6}$ | $<10^{-6}$ | $<10^{-6}$ | $<10^{-6}$ | $8.2\times10^{-5}$ | $5.0\times10^{-3}$ | 0.049 |
| S2 same population (single Rayleigh) | $<10^{-6}$ | $<10^{-6}$ | $1.8\times10^{-6}$ | $4.0\times10^{-5}$ | $1.9\times10^{-4}$ | 0.105 | 0.237 |
| S3 second population (adopted) | $2.2\times10^{-5}$ | $2.6\times10^{-3}$ | $7.7\times10^{-3}$ | 0.012 | $3.2\times10^{-3}$ | 0.053 | 0.102 |
| S3, $\sigma_{gross}$ = 8° | $<10^{-6}$ | $3.0\times10^{-4}$ | $1.4\times10^{-3}$ | $3.1\times10^{-3}$ | $5.2\times10^{-4}$ | 0.043 | 0.096 |
| S3, $\sigma_{gross}$ = 16° | $6.8\times10^{-4}$ | $7.4\times10^{-3}$ | 0.018 | 0.024 | 0.012 | 0.061 | 0.107 |
| S3, $\sigma_{gross}$ = 24° | $6.3\times10^{-3}$ | 0.014 | 0.033 | 0.039 | 0.036 | 0.069 | 0.111 |
| S3, π = 6.45% (cross-modal) | $1.1\times10^{-5}$ | $1.1\times10^{-3}$ | $3.3\times10^{-3}$ | $5.0\times10^{-3}$ | $1.4\times10^{-3}$ | 0.026 | 0.072 |
| S3, Student-t speed error | $3.2\times10^{-5}$ | $2.6\times10^{-3}$ | $7.8\times10^{-3}$ | 0.012 | $6.4\times10^{-3}$ | 0.058 | 0.089 |
| Tail share of S3 probability | >99% | >99% | >99% | >99% | 98% | 92% | 59% |

Table A2. P(bound) for the seven nominal events under alternative treatments of the 16.3° radiant discrepancy of 2019 MO. Under this absolute speed-error law the four strongest events keep the same order wherever their probabilities can be distinguished, but their P(bound) spans several orders of magnitude between scenarios. For those events more than 95% of the S3 probability comes from the gross component.

## A.3 Scaling of the speed error above the calibrated range

Table A3 compares the change needed to reach the parabolic boundary with the modern calibration in absolute and fractional terms. The largest modern residual is 2.15 km $s^{-1}$ in absolute terms and 18.6% of $v_g$ in fractional terms, both for 2019 MO. The fractional law uses the standard deviation of the modern fractional residuals, 5.8% of $v_g$; the direction model is unchanged. Required fractional changes are expressed relative to the CNEOS $v_g$.

| **Event** | **$v_g$ (km $s^{-1}$)** | **$\Delta v_g$ req. (km $s^{-1}$)** | **$\Delta v_g$ req. (%)** | **Δθ req. (°)** | **Core, absolute** | **Core, fractional** | **Core+tail, absolute** | **Core+tail, fractional** |
|---|---|---|---|---|---|---|---|---|
| 2014-01-08 (IM1) | 43.8 | 22.41 | 51.1 | 42.2 | $<10^{-6}$ | $<10^{-6}$ | $2.2\times10^{-5}$ | $4.1\times10^{-5}$ |
| 2026-04-01 (Polar-IM) | 68.2 | 10.76 | 15.8 | 23.0 | $<10^{-6}$ | $3.4\times10^{-3}$ | $2.6\times10^{-3}$ | 0.010 |
| 2017-03-09 (IM2) | 35.0 | 9.87 | 28.2 | 19.0 | $<10^{-6}$ | $2.5\times10^{-7}$ | $7.7\times10^{-3}$ | $8.8\times10^{-3}$ |
| 2022-07-28 (CNEOS-22) | 27.5 | 6.59 | 23.9 | 16.8 | $<10^{-6}$ | $1.9\times10^{-5}$ | 0.012 | 0.013 |
| 2009-04-10 | 15.7 | 2.89 | 18.4 | 28.3 | $8.2\times10^{-5}$ | $7.9\times10^{-4}$ | $3.2\times10^{-3}$ | $4.7\times10^{-3}$ |
| 2021-05-06 | 24.4 | 2.10 | 8.6 | 5.5 | $5.0\times10^{-3}$ | 0.073 | 0.053 | 0.113 |
| 2015-02-17 | 26.5 | 1.38 | 5.2 | 3.1 | 0.049 | 0.191 | 0.102 | 0.224 |

Table A3. Required speed and radiant changes for the seven nominal events, and P(bound) under absolute and fractional speed-error laws. Only Polar-IM changes qualitatively: its required speed change is five times the largest modern absolute residual but comparable to the largest modern fractional residual. Under the fractional law Polar-IM also overtakes 2017 March 9 in P(bound).

### A.4 Atmospheric deceleration or velocity definition?

If CNEOS reports a velocity at peak brightness while the references are pre-atmospheric, each residual should approximately equal minus the deceleration accumulated down to the reported height. We model that deceleration with a single-body drag law, $\ln(V_\infty/V) = \Gamma\,(A/m)\,\Sigma(h)/\cos z$, where $\Sigma(h) = p(h)/g$ is the vertical atmospheric column above height h in the U.S. Standard Atmosphere 1976, z is the zenith distance and A/m is the area-to-mass ratio of a sphere of bulk density 1500 kg m$^{-3}$. The effective coefficient $\Gamma = 1.06$, which absorbs drag, shape and fragmentation, is fixed as the median ratio over the 822 EFN824 fireballs with valid Vmax and photometric mass (Borovička et al. 2022). In that sample the fitted model has an in-sample association with the observed $V_\infty - V_{max}$ with Spearman $\rho = 0.68$, rising to 0.84 for the 165 fireballs with maximum brightness below 60 km (Figure A1a). The EFN median deceleration falls from 3.5 km s$^{-1}$ at 20–40 km to 0.35 km s$^{-1}$ above 90 km (Figure A1b). This is not an independent validation: the effective coefficient is calibrated on these same EFN data. The pressure prescription uses a geopotential-height approximation and an isothermal extension above 84.852 km, rather than the complete upper-atmosphere model.

For the CNEOS calibrators, the mass follows from the impact energy and reported speed, the zenith distance from the velocity vector and position, and the height is the reported peak-brightness altitude. The masses, $7\times10^2$ to $2\times10^5$ kg, lie far above the EFN range ($\leq$ 110 kg), and the model ignores the catastrophic fragmentation that often defines peak brightness for metre-scale bodies, so its event-level predictions are indicative only. Table A4 and Figure A1c summarise the test. The predicted mean deceleration (1.76 km s$^{-1}$) has the same sign and order of magnitude as the observed offset (1.00 ± 0.29 km s$^{-1}$), but it is larger, and no event-level association with the residual is detected. A significance test of the mean difference would require uncertainty in the model predictions, which is not available. In the high-quality sample across all epochs the mean residual is +0.25 km s$^{-1}$, because the pre-2018 network residuals are large and positive (Section 4.1), so the test is informative only for the modern subset. A definition or retrieval offset that is roughly constant across events would produce the same pattern. The test therefore does not identify the cause of the offset, which is why the stage shift is treated only as a sensitivity case. The independent speed checks of Section 7.4 give a mixed picture: the 2026 March 17 Ohio event is consistent with slowing, whereas for the 2026 August 14 Washington event the CNEOS speed exceeds the reported starting speed, the opposite of deceleration. We also searched both EFN catalogue releases listed in Table 1 (Borovička et al. 2022, 2025; 985 distinct fireballs, the second release being targeted at two meteor showers) for common events: none has a CNEOS counterpart, and the twelve fireballs with a CNEOS event within one hour are all more than 1,700 km away from it. The same holds for the GMN, SonotaCo, FRIPON and EDMOND catalogues, whose closest coincidence is 41.6 s and 357 km from the 2026 January 30 event. The model also uses a fixed effective area-to-mass ratio and a first-order conversion from reported-speed loss to geocentric-excess-speed loss. Its input mass is derived from CNEOS energy and speed, so the predicted loss is not statistically independent of the catalogue retrieval. The failure of this proxy cannot reject atmospheric deceleration in general or establish a constant retrieval offset. The no-match finding applies to the frozen CNEOS snapshot and the searched releases within the stated time and distance windows, not to all existing ground observations.

| Sample | n | Mean residual (km s$^{-1}$) | Mean predicted (km s$^{-1}$) | Slope (expected −1) | ρ predicted (p) | ρ height (p) |
|---|---|---|---|---|---|---|
| Modern high-quality | 9 | -1.00 | -1.76 | -0.06 ± 0.22 | -0.07 (0.86) | 0.38 (0.32) |
| High-quality, all epochs | 14 | 0.25 | -1.75 | -0.31 ± 0.47 | -0.11 (0.71) | 0.52 (0.05) |

Table A4. Test of the deceleration interpretation. "Mean predicted" is minus the mean drag-model deceleration to the reported CNEOS height. The slope and ρ predicted compare each event's residual with its own predicted deceleration; ρ height is the rank correlation between residual and reported height.

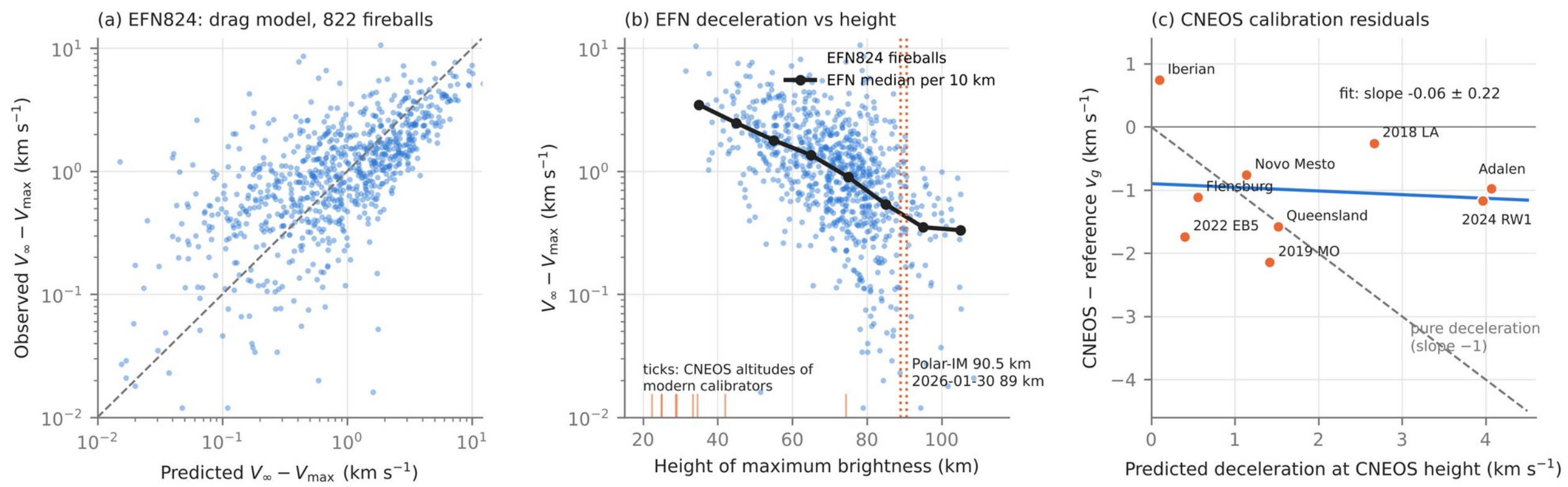

Figure A1. (a) Observed versus drag-model deceleration $V_\infty - V_{max}$ for 822 EFN824 fireballs; the dashed line is 1:1. (b) EFN deceleration against height of maximum brightness, with the median per 10 km; ticks mark the reported heights of the nine modern CNEOS calibrators and dotted lines the heights of Polar-IM and the 2026 January 30 event. (c) CNEOS-minus-reference speed residual of the modern calibrators against their predicted deceleration; pure deceleration would follow the dashed line of slope −1.